\documentclass[aps,prd,10pt,showpacs,amsmath,twocolumn,floatfix,amssymb,nofootinbib,longbibliography]{revtex4-2}
\usepackage{graphicx}
\usepackage{comment}
\usepackage[usenames]{color}
\usepackage{bm}
\usepackage{ifpdf}
\usepackage{floatrow}
\usepackage{makecell}
\usepackage{caption}
\usepackage{stackrel}
\usepackage{subcaption}
\usepackage{enumitem}

\usepackage[normalem]{ulem}
\usepackage[dvipsnames]{xcolor}
\usepackage[utf8]{inputenc}
\usepackage{hyperref}
\hypersetup{
  pdfnewwindow=true,      % links in new window
  colorlinks=true,        % false: boxed links; true: colored links
  linkcolor=PineGreen,    % color of internal links
  citecolor=PineGreen,    % color of links to bibliography
  filecolor=PineGreen,    % color of file links
  urlcolor=PineGreen      % color of external links
}
\newcommand{\be}{\begin{eqnarray}}
\newcommand{\ee}{\end{eqnarray}}

\usepackage{parskip}
\usepackage{anyfontsize}
\begin{document}

\title{Toward extracting scattering phase shift from integrated correlation functions V:  complex $\phi^4$ field model in $3+1$ dimensions}

\author{Peng~Guo}
\email{peng.guo@dsu.edu}

\affiliation{College of Arts and Sciences,  Dakota State University, Madison, SD 57042, USA}
\affiliation{Kavli Institute for Theoretical Physics, University of California, Santa Barbara, CA 93106, USA}
 
\author{Frank~X.~Lee}
\email{fxlee@gwu.edu}
\affiliation{Physics Department, The George Washington University, Washington, DC 20052, USA}

\author{Andrei~Alexandru}
\email{aalexan@gwu.edu}
\affiliation{Physics Department, The George Washington University, Washington, DC 20052, USA}

\date{\today}

\begin{abstract}
 In Ref.~\cite{Guo:2024zal} and associated studies, a relativistic finite-volume formalism in $1+1$ dimensions is proposed to extract infinite-volume scattering phaseshift.
 It is based on the difference of integrated correlation functions (ICF) rather than energy spectrum in the finite volume, and can be regarded as complementary to the well-known L\"{uscher} formalism. 
 In the present work,  the  formalism is further extended into $3+1$ dimensional spacetime. The aim is to explore and demonstrate the challenges in applying the formalism to more practical settings. Specifically, Monte Carlo simulations of a complex $\phi^4$ relativistic field model are carried out  in both 2+1 and 3+1 dimensions on lattices of varying sizes, and phaseshifts for the contact interaction are extracted from the formalism using modest computing resources.
\end{abstract}

\maketitle

\section{Introduction}\label{sec:intro}

Scattering is an indispensable tool in studying a wide range of dynamics, from atomic interactions in condensed matter physics to the strong interaction in quantum chromodynamics  (QCD).  Determination of scattering properties in such systems remains fundamental but challenging. In most cases, numerical simulations based on Monte Carlo evaluation of the path integral are performed by placing the system in a finite volume with periodic boundary conditions, which leads to quantized energy spectrum in the system. The energy spectrum  is then  connected  to the infinite volume scattering phaseshifts through quantization conditions.
         
 One such finite-volume formalism known as  the L\"uscher formula \cite{Luscher:1990ux}   has  proven   successful in a number of applications, especially in the meson sector, see e.g. Refs.~\cite{Aoki:2007rd,Feng:2010es,Lang:2011mn,Aoki:2011yj,Dudek:2012gj,Dudek:2012xn,Wilson:2014cna,Wilson:2015dqa,Dudek:2016cru,Beane:2007es, Detmold:2008fn, Horz:2019rrn,Guo:2020kph}. The  formalism  has also been   extended to beyond elastic scattering.  see e.g. Refs.~\cite{Rummukainen:1995vs,Christ:2005gi,Bernard:2008ax,He:2005ey,Lage:2009zv,Doring:2011vk,Guo:2012hv,Guo:2013vsa,Kreuzer:2008bi,Polejaeva:2012ut,Hansen:2014eka,Mai:2017bge,Mai:2018djl,Doring:2018xxx,Guo:2016fgl,Guo:2017ism,Guo:2017crd,Guo:2018xbv,Mai:2019fba,Guo:2018ibd,Guo:2019hih,Guo:2019ogp,Guo:2020wbl,Guo:2020kph,Guo:2020iep,Guo:2020ikh,Guo:2020spn,Guo:2021lhz,Guo:2021uig,Guo:2021qfu,Guo:2021hrf}.  
 However the L\"uscher formalism in few-nucleon systems still faces challenges such as the signal-to-noise (S/N)  ratio  of lattice correlation functions \cite{lepage1989analysis,DRISCHLER2021103888}  and the requirement of increasingly large number of interpolating operators  at large volumes \cite{Bulava:2019kbi}.

 These challenges motivated alternative approaches, such as the potential method of   HALQCD collaboration  \cite{PhysRevLett.99.022001,10.1143/PTP.123.89,PhysRevD.99.014514,ISHII2012437,AokiEPJA2013}, and the integrated correlation functions (ICF) formalism in \cite{Guo:2023ecc,Guo:2024zal,Guo:2024pvt,Guo:2025ngh,Guo:2025vgk}.  The latter relates  the difference between interacting and non-interacting integrated correlation functions  of two  particles in a finite volume  to the  infinite-volume phase shift through a weighted integral. It has the  advantage of  working directly with correlation functions, bypassing energy spectrum determination in traditional approaches.   Additional features include  the rapid  convergence at short Euclidean times  that makes it  potentially  a good candidate to overcome the S/N problem, and   free-by-construction from issues encountered at large volumes, such as increasingly dense energy spectrum and the extraction of low-lying states.

 So far,  discussions of the integrated correlation functions formalism are mostly limited to $1+1$ spacetime, including
 single channel non-relativistic dynamics~\cite{Guo:2023ecc},  relativistic dynamics~\cite{Guo:2024zal}, coupled-channel effect~\cite{Guo:2024pvt},  Coulomb corrections ~\cite{Guo:2025ngh}, and potential implementation on quantum computers~\cite{Guo:2025vgk}.  
 The aim of the present work is to extend it into $3+1$ spacetime dimensions. We explore and discuss some technical challenges in applying the formalism in higher dimensions. A relativistic complex $\phi^4$ field model on a finite lattice is used as a testbed to demonstrate in detail the feasibility of the integrated correlation functions approach via Monte Carlo simulations.
We will show that the formalism converges to the infinite volume limit and phase shifts are accessible in the more practical settings considered here.

The paper is organized as follows. A brief summary of integrated correlation function formalism is outlined in Sec.~\ref{sec:sumintegratedcorr},  with some technical details given in two appendices. Numerical test by Monte Carlo simulation of a complex $\phi^4$ lattice model in both $3+1$ and $2+1$ dimensions is presented in Sec.~\ref{sec:MCsimulation},   followed by summary in Sec.~\ref{sec:summary}.

\section{Relativistic integrated correlation function formalism in 3+1 dimensions}\label{sec:sumintegratedcorr}  
In this section, we outline the extension of relativistic integrated correlation function formalism for complex $\phi^4$ theory~\cite{Guo:2024zal} from $1+1$ to $3+1$ dimensions. More details can be found  in   Appendix \ref{scattsolutionsinfvol}.

In a periodic cubic box of size $L$, the correlation function that creates two particles at relative coordinate $ \mathbf{ r'} $ and time instant $0$ and then annihilates them at relative coordinate $\mathbf{ r}$ and time instant $t$ is defined by,
\begin{align}
C^{(2\phi)} (\mathbf{ r}t; \mathbf{ r'}0) & =  \langle 0 | \mathcal{T} \left [   \mathcal{O} (\mathbf{ r}, t )     \mathcal{O}^\dag (\mathbf{ r'} ,0)  \right ]  | 0 \rangle  , 
\end{align}
 where $\mathcal{T}$ stands for time-ordered operator.  The two-particle creation operator after projecting out center of mass motion (CM) in the rest frame is given by,
 \begin{equation}
    \mathcal{O}^\dag (\mathbf{ r} ,t) =  \frac{1}{\sqrt{2}} \frac{1 }{\sqrt{L^3}} \int_0^L d   \mathbf{ x}_2 \phi (\mathbf{ r}+\mathbf{ x}_2,t) \phi (\mathbf{ x}_2,t). \label{defreltwoparticleoperator}
 \end{equation}
The factor $1/\sqrt{2}$ takes into account the exchange symmetry of two distinguishable charged particles.   The $\phi$ field satisfies periodic boundary condition: $\phi (\mathbf{ x} + L \hat{\mathbf{ x}} , t) =\phi (\mathbf{ x}  , t) $, where $\hat{\mathbf{ x}}$ stands for a unit 3D spatial vector. 
The correlation function can be decomposed into momentum space and projected 
into the $A_1^+$  irreducible representation (irrep) of the cubic symmetry group,
\begin{equation}
C_{A_1^+}^{(2\phi)}(t) =  \frac{1}{L^3}  \sum_{\mathbf{ p} = \frac{2\pi \mathbf{ n}}{L}, \mathbf{ n} \in \mathbb{Z}^3} 2 \omega_p  \widetilde{C}_{A_1^+}^{(2\phi)}(\mathbf{ p}t; \mathbf{ p}0)  , \label{defintegratedCt}
\end{equation}
where $\omega_p = \sqrt{\mathbf{ p}^2 + m^2}$,  and $m$ is the renormalized mass of $\phi$ field.  The  momentum-space components are defined via double Fourier transforms,
\begin{equation}
 \widetilde{C}^{(2\phi)}(\mathbf{ p}t; \mathbf{ p'}0) =  \int_0^L d \mathbf{ r} d \mathbf{ r'}  e^{i \mathbf{ p} \cdot  \mathbf{ r}}   C^{(2\phi)}(\mathbf{ r}t; \mathbf{ r'}0)  e^{- i \mathbf{ p'} \cdot  \mathbf{ r'}}.
\end{equation}
The $A_1^+$ irrep projection is defined by double sums over group elements,
\begin{equation}
 \widetilde{C}_{A_1^+}^{(2\phi)}(\mathbf{ p}t; \mathbf{ p'}0) = \frac{1}{48^2} \sum_{ (g, g' ) \in \mathcal{G}}   \widetilde{C}^{(2\phi)}(g \mathbf{ p}t;  g' \mathbf{ p'}0)  ,  \label{twoctpA1proj}
\end{equation}
where $(\mathcal{G}, g)$ stand for the octahedral group $O_h$ and its elements, see e.g. Ref.~\cite{Doring:2018xxx,Cornwell:1997ke,Guo:2020kph}. 
The $O_{h}$ group has 10 irreps ($A_1^\pm$, $A_2^\pm$, $T_1^\pm$, $T_2^\pm$, $E^\pm$), including parity.
The $\phi^4$ theory is equivalent to a contact interaction theory, hence only $S$-wave contributes to the scattering dynamics, which only requires the $A_1^+$ irrep.

The integrated correlation function is related to    two-particle Green's function on the lattice by,
\begin{equation}
C^{(2\phi)}(\mathbf{ r} t; \mathbf{ r'}0)  =    i \int_{-\infty}^{\infty} \frac{d \lambda}{2\pi}  G^{(L)}  (\mathbf{r} ,  \mathbf{ r'}; \lambda)  e^{- i \lambda t}    .
\end{equation}
The two-particle Green's function is the solution of relativistic Dyson equation, with a contact interaction potential, 
\begin{equation}
    V(\mathbf{ r}) = V_0 \delta(\mathbf{ r}),
    \label{contact}
\end{equation}
 whose analytic solution is given by,
\begin{align}
& G^{(L)}(\mathbf{ r}, \mathbf{ r}'; E)   -    G^{(0, L)}(\mathbf{ r}- \mathbf{ r}'; E)   \nonumber \\
& =    \frac{G^{(0, L)}(\mathbf{ r}; E)     G^{(0, L)}(  \mathbf{ r}'; E)  }{\frac{1}{V_0} - G^{(0, L)}(\mathbf{ 0} ; E)  } .
\end{align}
The non-interacting two-particle Green's function  is defined by,
\begin{equation}
G^{(0, L)}(\mathbf{ r} ; E)   =      \frac{1}{L^3} \sum_{\mathbf{ q} = \frac{2\pi \mathbf{ n}}{L}, \mathbf{ n}\in \mathbb{Z}^3 } \frac{1   }{\omega_q}   \frac{   e^{i \mathbf{ q} \cdot \mathbf{ r}} }{  E^2 -  (2 \omega_q)^2  + i 0 }   .
\end{equation}
The difference of interacting and non-interacting integrated correlation functions is thus given in terms of Green's functions by,
\begin{align}
 & \triangle C_{A_1^+}^{(2\phi)}(t)   \nonumber \\
 &= \frac{1}{L^3} \sum_{\mathbf{ p} = \frac{2\pi \mathbf{ n}}{L}, \mathbf{ n} \in \mathbb{Z}^3 }   2 \omega_p  \left [   \widetilde{C}_{A_1^+}^{(2\phi)}(\mathbf{ p} t; \mathbf{ p} 0) -   \widetilde{C}_{A_1^+}^{(0, 2\phi)}(\mathbf{ p} t; \mathbf{ p} 0)  \right ] \nonumber \\
 &= i \int_{-\infty}^{\infty} \frac{d \lambda}{2\pi}   e^{- i \lambda t}   \nonumber \\
&  \times   \frac{1}{L^3} \sum_{\mathbf{ p} = \frac{2\pi \mathbf{ n}}{L}, \mathbf{ n} \in \mathbb{Z}^3}   2 \omega_p \left [\widetilde{G}_{A_1^+}^{(L)}(\mathbf{ p}, \mathbf{ p}; \lambda)-\widetilde{G}_{A_1^+}^{(0, L)}(\mathbf{ 0} ; \lambda)  \right ] .
\end{align}
 The $\widetilde{G}_{A_1^+}^{(L)}(\mathbf{ p}, \mathbf{ p}; \lambda)$ stands for the irrep projected Green's function in momentum space, which is defined in the same way as in  Eq.(\ref{twoctpA1proj}).
 
Using relativistic Krein's theorem relation in infinite volume (see Appendix \ref{relatFriedelformula}), 
    \begin{align} 
  &  \int_{-\infty}^\infty   \frac{d \mathbf{ p}}{(2\pi)^3}     \omega_p   \left [  \widetilde{G}^{(\infty)} (\mathbf{ p} , \mathbf{ p}; E)    -  \widetilde{G}^{(0, \infty)} (\mathbf{ 0}  ; E)  \right ]  \nonumber \\
  &  = -       \frac{1}{\pi} \int_{4 m^2}^\infty d s \frac{ \delta(\sqrt{s}) }{  (s - E^2 - i 0 )^2}    ,
\end{align}
the $S$-wave scattering phase shift of two scalar particles $\delta (E)$ is connected to  the difference of integrated correlation functions in its infinite volume limit through a weighted  integral,
\begin{equation}
 \triangle C_{A_1^+}^{(2\phi)}(t)   \stackrel{L \rightarrow \infty}{\rightarrow}    -     \frac{1}{\pi} \int_{2 m}^\infty d \epsilon   \delta( \epsilon)   \frac{d}{d \epsilon}  \left (  \frac{e^{ -  i  \epsilon t} }{ \epsilon}   \right ) . \label{mainresult}
\end{equation}
This relation is our main result (we will refer to it as the {\em ICF formula} for short).
Its Euclidean space-time version can be written in the form, 
\begin{equation}
 \triangle C_{A_1^+}^{(2\phi)}(\tau)   \stackrel{L \rightarrow \infty}{\rightarrow}     \frac{1}{\pi} \int_{2 m}^\infty d \epsilon   \delta( \epsilon)     \left ( \tau + \frac{1}{\epsilon}     \right )  \frac{e^{ -   \epsilon \tau} }{ \epsilon}   \label{mainresultEuclid}
\end{equation}
after analytic continuation ($t\to -i \tau$) and derivative.
This relation will be verified via Monte Carlo simulation in  the next section.
We should mention that although the above two formulas are presented in the context of a $\phi^4$ theory and contact interactions, they are more generally applicable to other field theory models, such as lattice QCD.
If projection to other irreps is required, a partial-wave expansion of the relations can be utilized, see Appendix \ref{scattsolutionsinfvol}.

\section{Monte Carlo Simulation test}\label{sec:MCsimulation}

We consider the complex $\phi^4$ action on a   $3+1$ Euclidean spacetime lattice, see e.g. Ref.~\cite{Guo:2018xbv,Guo:2024zal}, 
\begin{align} 
S_E  =& -   \kappa  \sum_{\mathbf{ x}, \tau, \hat{\mathbf{ n}}_\mathbf{ x}, \hat{n}_\tau  } \hat{\phi}^*(\mathbf{ x}, \tau) \hat{\phi}(\mathbf{ x}+ \hat{\mathbf{ n}}_\mathbf{ x}, \tau+\hat{n}_\tau )  + c.c. \nonumber \\
&+ (1-2 \lambda)  \sum_{\mathbf{ x}, \tau} | \hat{\phi}(\mathbf{ x},\tau) |^2 + \lambda \sum_{\mathbf{ x},\tau} |\hat{ \phi}(\mathbf{ x},\tau) |^4 , \label{Euclideanaction}
\end{align}
where  $(\mathbf{ x}, \tau)$ label the discrete coordinates of   Euclidean $L^3 \times T$  lattice sites:   $    (x,y,z) \in [0, L-1]$ and $\tau \in [0,T-1]$, and $(\hat{\mathbf{ n}}_\mathbf{ x}, \hat{n}_\tau )$ denote the unit vectors in direction $(\mathbf{ x},\tau)$. 
The $\phi$ field 
is rescaled on the lattice,  
\begin{equation}
\phi (\mathbf{ x},\tau)=  \sqrt{2\kappa } \hat{\phi} (\mathbf{ x},\tau),
\end{equation}
and it satisfies periodic boundary conditions in all directions.  The parameters $ (\kappa, \lambda)$ are related to bare mass $m_0$ and bare coupling constant  $g_0$ of interacting term $\frac{g_0}{4!} | \phi |^4$ by,
\begin{equation}
m_0^2 = \frac{1-2 \lambda}{\kappa}-8 \text{ and } g_0=\frac{6 \lambda}{\kappa^2}.  
\end{equation}
The lattice spacing $a$ is absorbed by rescaling in the parameters. If needed, the $a$ dependence can be restored by lattice units, $L\to L/a$, $m\to a\, m$,  {\it etc}. 
The $\phi^4$ term describes a contact interaction in the theory. 
The bare interaction strength $V_0$ in the contact potential model in 
Eq.\eqref{contact} can be adjusted by tuning $\lambda$.

Monte Carlo (MC) simulation is carried out by the Hybrid Monte Carlo (HMC) algorithm, see  details in Ref.~\cite{Guo:2018xbv}.  
 The single particle and two-particle  correlation functions are evaluated  respectively by path integrals,
\begin{equation}
 C_{lat}^{(\phi)}(\mathbf{ x} \tau, \mathbf{ x'} 0) =   \frac{   \int \mathcal{D} \phi \mathcal{D} \phi^\dag   \phi^\dag (\mathbf{ x}, \tau)  \phi ( \mathbf{ x'},  0)    e^{-S_E }  }{    \int \mathcal{D} \phi \mathcal{D} \phi^\dag     e^{-S_E }},
\end{equation}
and  
\begin{equation}
C_{lat}^{(2\phi)} (\mathbf{ r} \tau, \mathbf{ r' }0) =   \frac{   \int \mathcal{D} \phi \mathcal{D} \phi^\dag   \mathcal{O} (\mathbf{ r}, \tau)  \mathcal{O}^\dag (\mathbf{  r'},  0)    e^{-S_E }  }{    \int \mathcal{D} \phi \mathcal{D} \phi^\dag     e^{-S_E }},
\end{equation}
where subscript-$lat$ is added to denote the Monte Carlo simulation result, and the relative motion of two-particle interpolating operator  is defined in Eq.(\ref{reltwoparticleoperator}).

On the lattice, the {\em integrated} single-particle correlation function is defined by discrete sums over coordinates,
\begin{equation}
  C_{lat}^{(\phi)}( \tau)  = \sum_{ \mathbf{ x} \in [0, L-1]^3 } C_{lat}^{(\phi)}(\mathbf{ x} \tau, \mathbf{ x} 0) , \label{oneLatintegratedCt}
\end{equation}
and the {\em integrated} $A_1^+$-projected two-particle correlation function is defined by discrete sums over momentum,
\begin{equation}
C_{A_1^+,  lat}^{(2\phi)} (\tau) =  \frac{1}{L^3} \sum_{\mathbf{ p} = \frac{2\pi \mathbf{ n}}{L}}^{\mathbf{ n} \in   [ -\frac{L}{2} +1 , \frac{L}{2}]^3 }  2 \omega^{(lat)}_\mathbf{ p}  \widetilde{C}_{A_1^+, lat}^{(2\phi)}(\mathbf{ p} \tau, \mathbf{ p}0 ) , \label{twoLatintegratedCt}
\end{equation}
where the lattice dispersion relation is given by, 
 \begin{equation}
\omega^{(lat)}_\mathbf{ p} =\cosh^{-1} \left [ 3 + \cosh m -  \sum_{i = x, y, z} \cos \mathbf{ p}_i \right ]  .  \label{omegafiniteaandL}
\end{equation}
 The Fourier transform of  one or two particles correlation functions is defined by, 
\begin{align}
& \widetilde{C}_{lat}^{(\phi,2\phi)}( \mathbf{ p} \tau, \mathbf{ p'} 0) =  \nonumber \\
&\sum_{ \mathbf{ x}, \mathbf{ x'}  \in [0, L-1]^3 } e^{i \mathbf{ p} \cdot  \mathbf{ x}} C_{lat}^{(\phi,2\phi)}(\mathbf{ x} \tau, \mathbf{ x'} 0) e^{- i \mathbf{ p'} \cdot  \mathbf{ x'}} ,
\label{CtpFT}
\end{align}
where the discrete momenta in the three spatial dimensions are indicated by,
\begin{equation}
(\mathbf{ p} , \mathbf{ p'}  )= \frac{2\pi \mathbf{ n}}{L}, \  \mathbf{ n} \in  [ - \frac{L}{2} +1, \frac{L}{2}]^3.
\end{equation}

The most challenging part is to compute the integrated two-particle correlation functions in Eq.(\ref{twoLatintegratedCt}) which involves multiple nested 3D sums. 
We show in the following that this challenge can be overcome by Fourier transform.
If the $\phi$ field is transformed as,
\begin{equation}
 \widetilde{\phi} ( \mathbf{ p},\tau)   = \sum_{\mathbf{ x} \in [0, L-1]^3 }  e^{  i \mathbf{ p} \cdot  \mathbf{ x} }  \phi ( \mathbf{ x},\tau),
\end{equation}
then Eq.(\ref{twoLatintegratedCt}) can be expressed in momentum space as,
\begin{align}
& C_{A_1^+,  lat}^{(2\phi)} (\tau) = \label{CA12phi} \\
& \frac{1}{L^3} \sum_{\mathbf{ p} = \frac{2\pi \mathbf{ n}}{L}}^{\mathbf{ n} \in   [ -\frac{L}{2} +1 , \frac{L}{2}]^3 }  2 \omega^{(lat)}_\mathbf{ p}  \langle 0 |     \widetilde{\mathcal{O}}_{A_1^+} (\mathbf{ p}, \tau )     \widetilde{\mathcal{O}}_{A_1^+}^\dag (\mathbf{ p} ,0)   | 0 \rangle . \nonumber 
\end{align}
The irrep-projected two-particle correlation function is defined by, 
\begin{equation}
   \widetilde{\mathcal{O}}_{A_1^+}^\dag (\mathbf{ p}, \tau )  = \frac{1}{48} \sum_{g \in \mathcal{G}}    \widetilde{\mathcal{O}}^\dag (g \mathbf{ p}, \tau ),
\end{equation}
where
 \begin{equation}
   \widetilde{\mathcal{O}}^\dag (\mathbf{ p}, \tau )  =  \frac{1}{\sqrt{2}} \frac{1 }{\sqrt{L^3}}     \widetilde{\phi} (\mathbf{ p}, \tau )       \widetilde{\phi} ( - \mathbf{ p}, \tau )    .
\end{equation}

Using the identity,
\begin{equation}
\sum_{\mathbf{ p} = \frac{2\pi \mathbf{ n}}{L}}^{\mathbf{ n} \in  \left [ - \frac{L}{2} + 1 , \frac{L}{2} \right ]^3 }    =  \sum_{\mathbf{ p}_0  = \frac{2\pi \mathbf{ n}}{L}}^{\mathbf{ n} \in  \left [ 0 , \frac{L}{2}  \right ]^3 } \frac{ \vartheta (\mathbf{ p}_0) }{\sigma_{\mathbf{ p}_x}  \sigma_{\mathbf{ p}_y}  \sigma_{\mathbf{ p}_z} }  ,\label{p0identity}
\end{equation}
where the symmetry factor $\sigma_{\mathbf{ p}_i }$ is defined by
\begin{equation}
\sigma_{\mathbf{ p}_i } = \begin{cases}  2, & \text{if} \  \ \frac{L}{2\pi} {\mathbf{ p}_i }  = \frac{1}{2}  , \\ 1, & \text{otherwise}, \end{cases}
\end{equation}
Eq.(\ref{CA12phi}) can be further reduced to
\begin{align}
 C_{A_1^+,  lat}^{(2\phi)} (\tau) & =  \frac{1}{L^3} \sum_{\mathbf{ p}_0 = \frac{2\pi \mathbf{ n}_0}{L}}^{\mathbf{ n}_0 \in   [ 0, \frac{L}{2}]^3 }  \frac{ \vartheta (\mathbf{ p}_0) }{\sigma_{\mathbf{ p}_x}  \sigma_{\mathbf{ p}_y}  \sigma_{\mathbf{ p}_z} }   \nonumber \\
& \times 2 \omega^{(lat)}_{\mathbf{ p}_0 }  \langle 0 |    \widetilde{\mathcal{O}}_{A_1^+} (\mathbf{ p}_0, \tau )     \widetilde{\mathcal{O}}_{A_1^+}^\dag (\mathbf{ p}_0 ,0)   | 0 \rangle .
\end{align}
 The $\mathbf{ p}_0$ denote a single reference vector in cubic symmetry group.  It is the set of momenta represented by $\mathbf{ p}_0$ via $\mathbf{ p} = g \mathbf{ p}_0 (g \in \mathcal{G})$. 
 The $\vartheta(\mathbf{ p}_0)$ is the multiplicity of distinct momenta within the set represented by $\mathbf{ p}_0$, see e.g. Ref.~\cite{Doring:2018xxx,Cornwell:1997ke,Guo:2020kph} and also see some details in Appendix \ref{modelsetup}.
 
In summary, computation of the two-particle  integrated correlation function can be greatly accelerated by employing Fourier transform of the $\phi$ field. We pre-compute the transform by using a Fast Fourier Transform software package called {\em fftw3} and store it in memory. As a result, the cost of computing integrated two-particle correlation function is on par with that of single-particle ones.

\subsection{Exact solutions for non-interacting case: $\lambda =0$}

In the absence of interactions, {\it i.e.}, setting $\lambda =0$ in Euclidean action in Eq.(\ref{Euclideanaction}), analytic expressions of correlation functions  can be obtained, see e.g. Ref.~\cite{Guo:2024zal}.  This fact can be used to numerically check some basic properties of the Monte Carlo simulation.  
First, single-particle correlation in $3+1$ dimensions is given by,
\begin{equation}
    C^{(0, \phi)}(\mathbf{ x} \tau, \mathbf{ x'} 0)     =   \frac{1}{L^3} \sum_{\mathbf{ k} = \frac{2\pi \mathbf{ n}}{L} }^{\mathbf{ n}  \in   [- \frac{L}{2} +1, \frac{L}{2}]^3 }    e^{i \mathbf{ k} \cdot  (\mathbf{ x}-\mathbf{ x'}) }  G_{\phi}  (\mathbf{ k}, \tau),
\end{equation}
and two-particle correlation function by,
\begin{align}
 &   C^{(0, 2\phi)}(\mathbf{ r} \tau, \mathbf{ r'} 0)  \\
 &  =\frac{1}{L^3} \sum_{\mathbf{ k} = \frac{2\pi \mathbf{ n}}{L} }^{\mathbf{ n} \in  [- \frac{L}{2} +1, \frac{L}{2}]^3  }  \cos (\mathbf{ k} \cdot  \mathbf{ r}) \cos ( \mathbf{ k} \cdot  \mathbf{ r'})     \left [  G_{\phi}  (\mathbf{ k}, \tau)   \right ]^2  .  \nonumber 
\end{align}
The Green's function in the above expressions is defined by,
\begin{align}
& G_{\phi }  (\mathbf{ k}, \tau) =  \label{Gphidef} \\
&  \frac{1}{T} \sum_{\omega = \frac{2\pi n}{T}}^{n \in [0,T-1]}      \frac{ e^{i \omega \tau   }   }{ \sum_{ i = x, y, z} (2  - 2   \cos  \mathbf{ k}_i  ) -2  \cos \omega        + 2 \cosh m    } . \nonumber
\end{align}

Next, using the Fourier transform in Eq.\eqref{CtpFT}, the integrated one-particle correlation function is reduced to a single sum on the Green's function,
 \begin{align}
C^{(0, \phi)}(  \tau)  &  =   \frac{1}{L^3} \sum_{\mathbf{ p} = \frac{2\pi \mathbf{ n}}{L}}^{\mathbf{ n} \in   [- \frac{L}{2} +1, \frac{L}{2}]^3 } \widetilde{C}^{(0, \phi)}( \mathbf{ p} \tau, \mathbf{ p} 0)   \nonumber \\
&=    \sum_{\mathbf{ p} = \frac{2\pi \mathbf{ n}}{L}}^{\mathbf{ n} \in   [- \frac{L}{2} +1, \frac{L}{2}]^3 }  G_{ \phi }  (\mathbf{ p}, \tau)    . \label{integratedCphit}
\end{align}
The integrated two-particle correlation function is reduced to double sums over momenta,
\begin{align}
   C_{A_1^+}^{(0, 2 \phi)}(  \tau)  & =   \frac{1}{L^3} \sum_{\mathbf{ p} = \frac{2\pi \mathbf{ n}}{L}}^{\mathbf{ n} \in   [ -\frac{L}{2} +1 , \frac{L}{2}]^3 }  2 \omega^{(lat)}_\mathbf{ p}  \widetilde{C}_{A_1^+}^{(0, 2\phi)}(\mathbf{ p} \tau, \mathbf{ p}0 )   \nonumber \\
   & =     \sum_{ (\mathbf{ p} , \mathbf{ k}) \in \frac{2\pi \mathbf{ n}}{L}}^{\mathbf{ n} \in   [ -\frac{L}{2} +1 , \frac{L}{2}]^3 }  2 \omega^{(lat)}_\mathbf{ p}    \sigma^{(lat)}_{\mathbf{ p}, \mathbf{ k}}  \left [  G_{\phi}  (\mathbf{ k}, \tau)   \right ]^2      , \label{integratedCtwophit}
\end{align}
where the weighting factor 
 is defined by,
 \begin{align}
 \sigma^{(lat)}_{\mathbf{ p}, \mathbf{ k}}  & =   \left | \frac{1}{48} \sum_{g}  \frac{1}{L^3} \int d \mathbf{ r} e^{i g \mathbf{ p} \cdot \mathbf{ r}} \cos (\mathbf{ k} \cdot  \mathbf{ r}) \right |^2     \\
  & = \begin{cases}   
  \frac{ \sigma^2_{ \mathbf{ p}_{x} }  \sigma^2_{ \mathbf{ p}_{y}  }  \sigma^2_{ \mathbf{ p}_{z} } }{\vartheta^2 (\mathbf{ p}_0)}      , & \text{if} \ (\mathbf{ k}, \mathbf{ p}) \in g \mathbf{ p}_0 , \ \text{where}  \ g \in \mathcal{G},  \\   0, & \text{otherwise}. \nonumber
  \end{cases}
\end{align}
Using the identity in Eq.(\ref{p0identity}) again,  Eq.(\ref{integratedCtwophit}) can be further reduced to a single sum,
\begin{equation}
   C_{A_1^+}^{(0, 2 \phi)}(  \tau)       =     \sum_{ \mathbf{ p}_0 = \frac{2\pi \mathbf{ n}_0 }{L} }^{\mathbf{ n}_0  \in   [ 0, \frac{L}{2}]^3}  2 \omega^{(lat)}_{\mathbf{ p}_0}       \left [  G_{\phi}  (\mathbf{ p}_0, \tau)   \right ]^2      .  \label{ct0A1analytic}
\end{equation}

In the continuum limit of $a \rightarrow 0$ and $T \rightarrow \infty$,  the  integrated correlation functions  approach  known forms for free particles,
\begin{equation}
C^{(0, \phi)}(  \tau)    \stackrel[a\rightarrow 0]{T \rightarrow \infty}{ \rightarrow }     \sum_{ \mathbf{ p} = \frac{2\pi \mathbf{ n}}{L}}^{\mathbf{ n} \in \mathbb{Z}^3}    \frac{e^{- \omega_p \tau}  }{2 \omega_p } ,  
\end{equation}
and
\begin{equation}
   C_{A_1^+}^{(0, 2 \phi)}(  \tau)     \stackrel[a\rightarrow 0]{T \rightarrow \infty}{ \rightarrow }    \sum_{ \mathbf{ p}_0  }  \frac{e^{- 2 \omega_{p_0} \tau}  }{ 2 \omega_{p_0}},
\end{equation}
  where $\omega_q = \sqrt{\mathbf{ q}^2 + m^2}$ for both $\mathbf{ q}= \mathbf{ p}$ and $\mathbf{ q}= \mathbf{ p}_0$.
 It serves as another consistency check of the formalism.

  \begin{figure}
\includegraphics[width=0.99\textwidth]{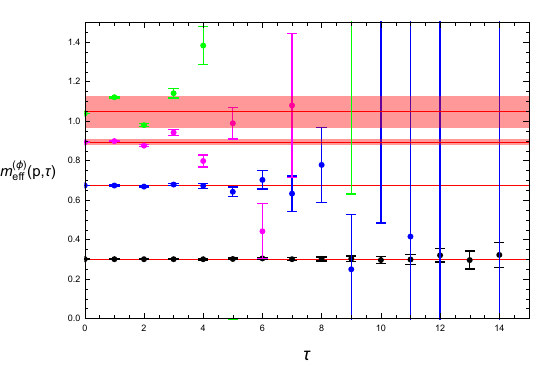}
 \caption{Non-interacting single-particle effective mass from Monte Carlo data of $m^{(\phi)}_{eff}$ in Eq.\eqref{emass} at the lowest three momenta $\mathbf{ p}=2\pi \mathbf{n}/L$  with $\mathbf{n}= (0,0,0)$ (black), $(1,0,0)$ (blue), $(1,1,0)$ (purple), and $(1,1,1)$ (green). The fit results (red bands) are also indicated.  
 \label{onemassL10plot} }
 \end{figure}

  \begin{figure}
\includegraphics[width=0.99\textwidth]{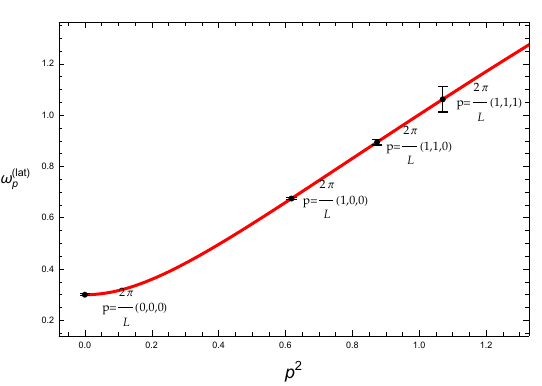}
 \caption{Lattice dispersion relation defined in Eq.(\ref{omegafiniteaandL}) is compared with the simulation results measured in Fig.~\ref{onemassL10plot}.  \label{dispersonL10plot} }
 \end{figure}

  \begin{figure}
\includegraphics[width=0.99\textwidth]{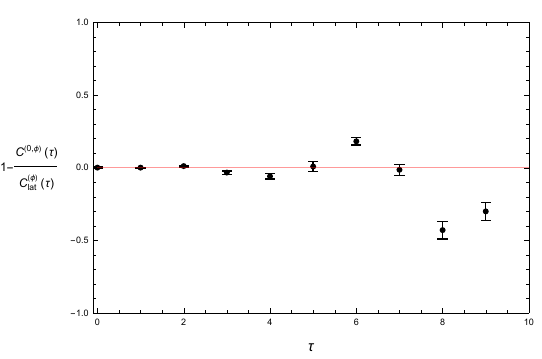}
 \caption{Comparison of lattice results for non-interacting single-particle integrated correlation function against the exact expression in Eq.(\ref{integratedCphit}).  The parameters are: $T=60$, $L=10$, $\kappa=0.0618$ and $\lambda=0$.  \label{onefreeL10diffCtplot} }
 \end{figure}

   \begin{figure}
\includegraphics[width=0.99\textwidth]{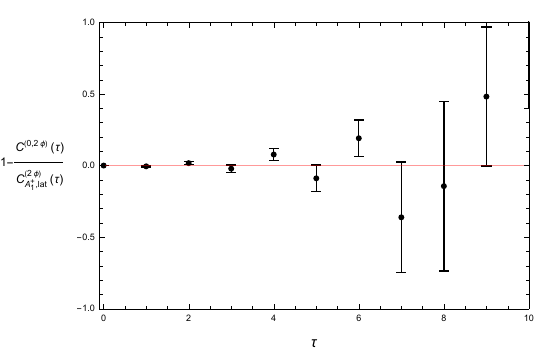}
 \caption{Similar to Fig.~\ref{onefreeL10diffCtplot}, but for the free two-particle integrated correlation function in Eq.(\ref{integratedCtwophit}).  \label{twofreeLdiffCtplot} }
 \end{figure} 

Monte Carlo simulations are performed for non-interacting charged scalar particles with the following choice of parameters:  $\kappa =0.0618$, and $\lambda  = 0$,  fixed temporal extent $T=60$, spatial extent $L=10$, and 200k configurations. 
The mass of the single particle is measured by fitting single-particle correlation function $ \widetilde{C}^{(0,\phi)}( \mathbf{ p} \tau, \mathbf{ p} 0 ) $ at zero momentum ($\mathbf{ p}=0$). We find $m = 0.299 \pm 0.004$. Note that $m$ is the only free parameter in the single-particle  correlation function which appears via Eq.\eqref{Gphidef}.

The effective mass defined by,
\begin{equation}
m^{(\phi,2\phi)}_{eff} (\mathbf{ p},\tau) = \ln \frac{ \widetilde{C}^{(\phi,2\phi)}_{lat}( \mathbf{ p} \tau, \mathbf{ p} 0 ) }{ \widetilde{C}^{(\phi,2\phi)}_{lat}( \mathbf{ p} (\tau+1), \mathbf{ p} 0)   },
\label{emass}
\end{equation}  
is plotted in Fig.~\ref{onemassL10plot} for free single-particle correlation functions at different momenta. The signal is good at zero momentum, but deteriorates increasingly at non-zero momenta.

The lattice dispersion relation in Eq.(\ref{omegafiniteaandL}) is checked in Fig.~\ref{dispersonL10plot} using lattice momentum,
\begin{equation}
p^2= 4  \sum_{i=x,y,z} \sin^2 (\frac{\mathbf{ p}_i}{2}),    
\end{equation}
and  the measured mass and energies in Fig.~\ref{onemassL10plot}. We see that the relation is well satisfied, albeit with larger errors at higher momenta.

Next, in Fig.~\ref{onefreeL10diffCtplot} we check the lattice results against the exact single-particle integrated correlation function in Eq.(\ref{integratedCphit}) by plotting $ 1-  C^{(0, \phi)} (  \tau)/C^{(\phi)}_{lat}(  \tau)$ 
which should approach zero for perfect agreement.  We see good agreement at small times.
The non-interacting two-particle integrated function is checked in Fig.~\ref{twofreeLdiffCtplot} against Eq.(\ref{integratedCtwophit}). Similar agreement is observed at small times, but worse signal at large times.
The two-particle correlation function takes as input the mass $m$ fitted from the single-particle correlation function.

\subsection{Phase shift analysis in interacting cases: $\lambda \neq 0$}

Monte Carlo simulations for interacting charged scalar particles are carried out  with the following choice of parameters:  $\kappa =0.0666$, and $\lambda  = 0.03$,  $T=60$,  and  spatial  extents ranging from $L=4$ to 28. We use statistics on the order of 200k measurements for each case.

  \begin{figure}
\includegraphics[width=0.99\textwidth]{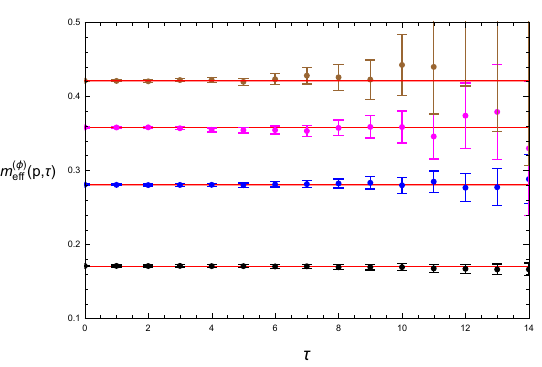}
 \caption{Interacting single-particle effective mass plot: Monte Carlo data (colored points with error bars) vs. fit results (red bands). The color coding for momentum $\mathbf{ p}=\frac{2\pi}{L } \mathbf{n}$ with $\mathbf{n}=(0,0,0)$ (black), $(1,0,0)$ (blue), $(1,1,0)$ (purple), and  $(1,1,1)$ (brown).  The parameters are: $T=60$, $L=28$, $\kappa=0.0666$ and $\lambda=0.03$.  \label{oneintmassL28plot} }
 \end{figure}

 \begin{figure}
\includegraphics[width=0.99\textwidth]{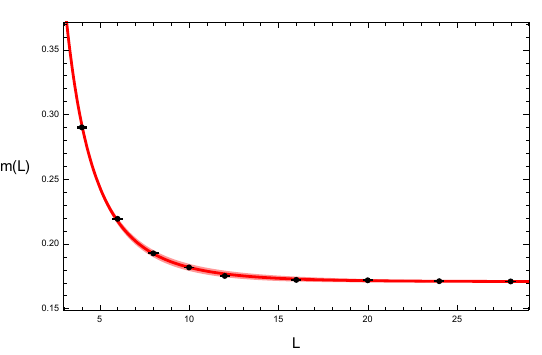}
 \caption{Lattice size dependence of interacting single-particle mass in Eq.\eqref{mL}  is fitted to MC data using the parameters $T=60$, $\kappa=0.06255$ and $\lambda=0.005$.  \label{onemassallLplot} }
 \end{figure}

  \begin{figure}
\includegraphics[width=0.99\textwidth]{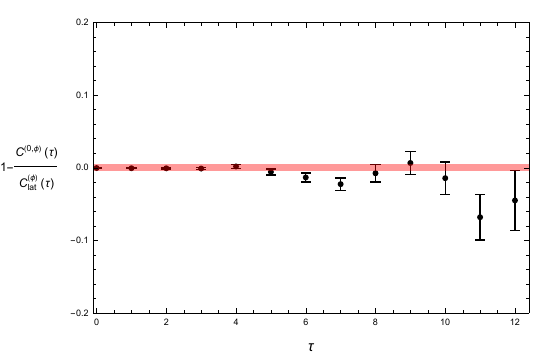}
 \caption{ Comparison of lattice  vs. exact single-particle integrated  correlation functions for the interacting case in $3+1$ dimensions.  The parameters are: $T=60$, $L=28$, $\kappa=0.0666$ and $\lambda=0.03$.  \label{oneintL28diffCtplot} }
 \end{figure}

  \begin{figure*}
\includegraphics[width=0.89\textwidth]{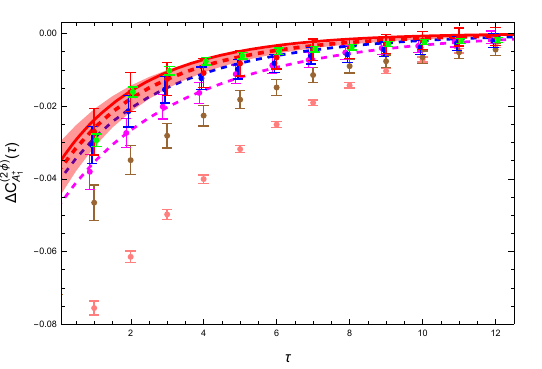}
 \caption{Convergence of the relation in Eq.\eqref{mainresultEuclid} in $3+1$ dimensions.
 The left-hand-side $ \triangle C_{A_1^+}^{(2\phi)}(  \tau)$ from MC simulation is shown as data points on lattices of size $L=12$ (pink), $16$ (brown), $20$ (magenta), $24$ (blue) and $28$ (red). The data points for $L=20, 24$ are  slightly shifted horizontally for better viewing. The perturbation result is color-matched for $L=20$ (dashed magenta), $24$ (dashed blue), and  $28$ (dashed red). The red band is uncertainty propagated from fitting $L=24, 28$ to perturbation result to obtain $V_R$.   The infinite volume limit is the solid red curve.  The fixed parameters are: $T=60$, $\kappa=0.0666$ and $\lambda=0.03$.  (The green points are   high statistic results with about one million configurations  for $L=28$.)\label{diffCtplot} }
 \end{figure*}

  \begin{figure}
\includegraphics[width=0.99\textwidth]{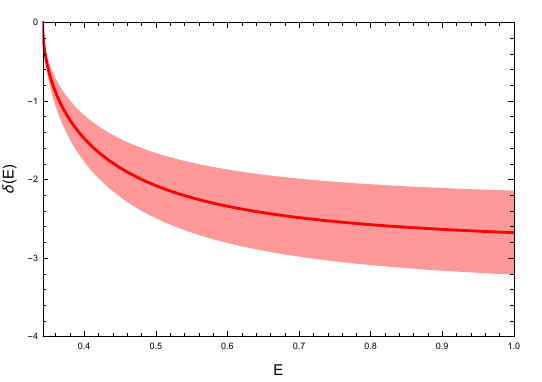}
 \caption{Phase shift (in degrees) as a function of energy extracted from Eq.\eqref{phaseshift4D} in $3+1$ dimensions. \label{phaseshiftplot} }
 \end{figure}

The single-particle mass  is measured by fitting momentum-projected single-particle correlation function $ \widetilde{C}^{(0, \phi)}( \mathbf{ p} \tau, \mathbf{ p} 0) $   with $\mathbf{ p}=0$.
Fig.~\ref{oneintmassL28plot} shows an example of single-particle effective mass at several momenta. The data quality is fairly good at all momenta considered. Interestingly, the interacting signal is better than that in the non-interacting case in Fig~\ref{onemassL10plot}. 

In Fig.~\ref{onemassallLplot} we fit our data to the well-known lattice size dependence of single-particle mass,  see e.g. Ref.~\cite{Rummukainen:1995vs},
\begin{equation}
m(L) =m + \frac{c}{\sqrt{L^3}} e^{- m L} ,
\label{mL}
\end{equation}
 and obtain $m = 0.17084 \pm 0.0003$ and $c=1.92\pm 0.02$. It is smaller than the one ($m=0.299$) in the non-interacting case. This mass in the interacting case is used as input in two-particle correlation functions.
 
Fig.~\ref{oneintL28diffCtplot} shows the comparison plot of single-particle integrated correlation function $ 1-  C^{(0, \phi )}(  \tau)/C^{(\phi )}_{lat}(  \tau)$.  This figure differs from Fig.~\ref{onefreeL10diffCtplot} mainly in that one is interacting case ($\kappa=0.0666$, $\lambda=0.03$, $m = 0.17084$), and the other non-interacting ($\kappa=0.0618$, $\lambda=0$, $m = 0.299$).
 
To verify the main relation in Eq.\eqref{mainresultEuclid},
first, the difference of integrated two-particle correlation functions between lattice MC data and non-interacting analytic expression (left-hand-side), 
\begin{equation}
\triangle C_{A_1^+}^{(2 \phi)}(  \tau) = C^{(2 \phi)}_{A_1^+, lat}(  \tau)  - C^{(0,2 \phi)}_{A_1^+}(  \tau) ,
\end{equation}
 is computed on various lattices of size $L=12, 16,20,24,28$.
Here the interacting term is given in Eq.\eqref{twoLatintegratedCt}, and the 
 non-interacting term in Eq.(\ref{integratedCtwophit}).
 Next, we employ perturbation result for the right-hand-side, 
 \begin{equation}
\triangle C^{(2 \phi)}_{pert} (  \tau)  = -  \frac{V_R(0) }{L^3} \sum_{\mathbf{ p} = \frac{2\pi \mathbf{ n}}{L}}^{\mathbf{ n} \in    \mathbb{Z}^3 }    \frac{\tau + \frac{1}{2\omega_\mathbf{ p}}}{(2\omega_\mathbf{ p})^3} e^{- 2 \omega_\mathbf{ p} \tau},
\label{pert}
\end{equation}
to fit the difference on the left-hand-side to obtain the renormalized coupling strength $V_R(0)$ in the contact interaction. The perturbation expression is $L$ dependent. 
Details of the perturbation expansion are outlined in Appendix~\ref{perturbationresult}.

Once the renormalized coupling strength $V_R(0)$ is extracted, the $S$-wave phase shift can be obtained by the quantization condition, see Appendix \ref{scattsolutionsinfvol},
\begin{equation}
\delta (E) = \cot^{-1}  \left [ - \frac{\frac{1}{V_R(0)}  -  Re \left [ G(E) \right ]+ G(0) }{\rho(E)} \right ] , \label{phaseshift4D}
\end{equation}
 where
 \begin{equation}
  G(E)  =    \frac{\rho(E)}{ \pi }  \ln \frac{\rho(E) +1}{ \rho(E) -1}   ,
\end{equation}
and
  \begin{equation} 
\rho(E) =   \frac{ 1 }{16 \pi }      \sqrt{1 -  \frac{ 4 m^2 }{E^2} } .  
\end{equation}
Finally, the result is compared with the infinite volume limit on the right-hand-side of Eq.\eqref{mainresultEuclid} with input from  $V_R(0)$ determined in perturbation theory and  phaseshift expression in Eq.\eqref{phaseshift4D}.

The result from the above-mentioned analysis strategy is summarized in Fig.~\ref{diffCtplot}. 
Although the complex $\phi^4$ theory is a relatively simple field model,  its MC simulation in $3+1$ dimensions is still a demanding numerical undertaking, especially on large lattices. With configurations on the order of 200k, the overall quality of the signal in Fig.~\ref{diffCtplot} is not ideal. Specifically, the  $L=28$  signal for $\triangle C_{A_1^+}^{(2\phi)}( \tau ) $ fluctuates for $\tau >4$, which is also observed in the single-particle correlation function in  Fig.~\ref{oneintL28diffCtplot}.  Nonetheless, the convergence of the relation to the infinite volume limit is clearly visible as $L$ is increased, despite the relatively large uncertainties in the data at $L=24, 28$.  Fitting the signal from these two lattices with perturbation theory, we can get a estimate of the renormalized contact interaction potential strength,
 $ 
  V_R(0) =  2.5 \pm 0.5. 
 $
Using this value, the phase shift can be extracted via Eq.(\ref{phaseshift4D}), as illustrated in Fig.~\ref{phaseshiftplot}. The uncertainty band is a direct consequence of the large uncertainties in the two-particle ICF signal. As the signal improves, the band is expected to shrink accordingly, as confirmed next.

\subsection{Sanity check in $2+1$ dimensions  }

To lend support for the extension from $1+1$ to $3+1$ dimensions, we consider the in-between case of $2+1$ dimensions in this section.
The reduction of the formalism from $3+1$ dimensions presented in previous sections down to $2+1$ dimensions is fairly straightforward. 
The main relation in Eq.\eqref{mainresultEuclid} takes the same form.
All the definitions of one- and two-particle correlation functions remain the same. The primary difference lies in the expression of $S$-wave phase shift, and symmetry group projections in $2+1$ dimensions.

The scattering amplitude and $S$-wave phase shift for the complex $\phi^4$ theory  in $2+1$ dimensions can be derived in a similar way to $3+1$, as detailed in the appendices. Here we only point the crucial differences for the convergence test. The free particle Green's function at origin  is now given  by,
   \begin{equation} 
G^{(0, \infty)} (\mathbf{ 0}; E)    = \frac{ 1}{8\pi}  \int_{4 m^2}^\infty  \frac{d s' }{ s' } \frac{\sqrt{s'} }{E^2 -  s' + i 0 },
\end{equation}
whose analytic expression is free of ultraviolet divergence,
\begin{equation}
 G^{(0, \infty)} (\mathbf{ 0}; E)  =  G(E)   =     \frac{1}{8\pi E}  \ln \frac{ \frac{2m}{E} -1}{ \frac{2m}{E} +1}  .
\end{equation}
Hence  the  coupling strength of the contact interaction   in $2+1$ dimensions  does not require renormalization. The scattering amplitude is given simply by, 
\begin{equation}
t(E)   = - \frac{1}{ \frac{1}{ V_0 }  - G(E) }  = \frac{1}{\rho(E)} \frac{1}{\cot \delta(E) - i },  
\end{equation}
where the phase space factor is defined by
\begin{align}
\rho(E) = \frac{1}{8 E} .
\end{align}
The analytic expression of $S$-wave phase shift is related to $V_0$ by,
\begin{equation}
\delta (E) = \cot^{-1}  \left [  \frac{\frac{1}{V_0}  -  Re \left [ G(E) \right ]  }{  Im \left [ G(E) \right ]} \right ] .
\label{phase2}
\end{equation}

The symmetry group is reduced from that of a cube (octahedral group $O_h$ with 48 elements) in 3+1 dimensions to that of a square (dihedral group $D_{4}$ with 8 elements) in $2+1$ dimensions. 
The $D_{4}$ group has five irreps ($A_1$, $A_2$, $B_1$, $B_2$, $E$), not counting parity.
The projection to the $A_1$ irrep corresponds to the S-wave in $2+1$ dimensions,
\begin{equation}
 \widetilde{C}_{A_1}^{(2\phi)}(\mathbf{ p}t; \mathbf{ p'}0) = \frac{1}{8^2} \sum_{ (g, g' ) \in \mathcal{G}}   \widetilde{C}^{(2\phi)}(g \mathbf{ p}t;  g' \mathbf{ p'}0)  , 
\end{equation}
 where  the $(\mathcal{G}, g)$ now refer to the $D_{4}$ group  and its elements.

   \begin{figure}[b]
\includegraphics[width=0.99\textwidth]{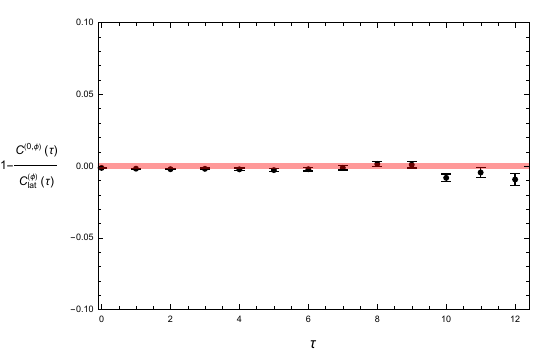}
 \caption{Comparison of lattice  vs. exact single-particle integrated  correlation functions in $2+1$ dimensions.  The parameters are: $T=60$, $L=28$, $\kappa=0.0875$ and $\lambda=0.02$. Its counterpart in $3+1$ dimensions is Fig.~\ref{oneintL28diffCtplot}. \label{oneintL28diffCtplot3D} }
 \end{figure} 

  \begin{figure*}
\includegraphics[width=0.89\textwidth]{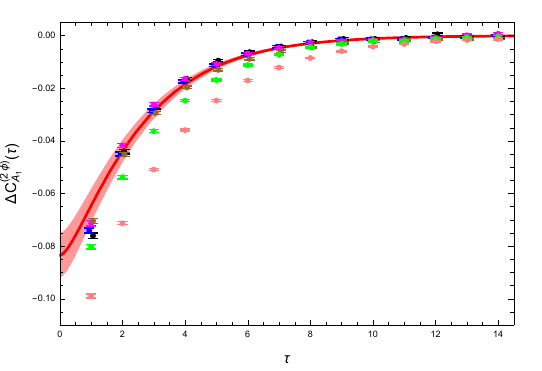}
 \caption{Convergence of the relation in Eq.\eqref{mainresultEuclid} in $2+1$ dimensions.
 The left-hand-side $ \triangle C_{A_1}^{(2\phi)}(  \tau)$ from MC simulation is shown as data points on lattices of size $L=12$ (pink), $16$ (green), $20$ (brown),  $24$ (blue), $28$ (magenta), and $32$ (black). The infinite volume limit is the solid red curve.  Other parameters are: $T=60$, $\kappa=0.0875$ and $\lambda=0.02$. \label{diffCtplot3D} }
 \end{figure*} 

  \begin{figure}[t]
\includegraphics[width=0.99\textwidth]{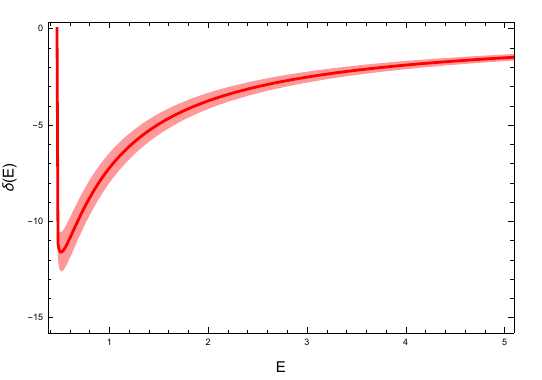}
 \caption{Phase shift (in degrees) extracted from Eq.\eqref{phase2}  in $2+1$ dimensions.  \label{phaseshiftplot3D} }
 \end{figure}

The computational demand in $2+1$ dimensions scales down by about a factor of $L$ relative to $3+1$ dimensions.  Consequently, we can perform high-statistics  Monte Carlo simulations, typically a million configurations for each case.  For interacting charged scalar particles we use the following parameters:  $\kappa =0.0875$, and $\lambda  = 0.02$.  The temporal extent of the lattice  is  again fixed at $T=60$, and multiple lattices up to $L=32$ are considered. The single-particle mass is found with relatively high precision: $m=0.2353 \pm 0.0003$. The single-particle integrated  correlation function from MC simulation matches well with the exact result up to $\tau=9$, as shown in Fig.~\ref{oneintL28diffCtplot3D}.  The convergence test is given in Fig.~\ref{diffCtplot3D} for $L=12,16,20,24,28 ,32$. 
We see that, with high statistics, the finite-volume lattice results for  $L=24,28 ,32$  demonstrate  excellent convergence and match the infinite volume result fairly well overall. Small deviations are observed at time $\tau=1$ between the lattice data and its infinite volume limit. The cause of the deviation is not clear at the moment. It could be an indication of inelastic contribution or the noisy  signal of Monte Carlo simulation at high energies. 
Nonetheless, we obtain  $V_0 =1.1 \pm 0.1 $ for the coupling strength in the contact interaction potential directly by fitting the right-hand-side of   Eq.\eqref{mainresultEuclid} using data on $L=24,28,32$ and Eq.\eqref{phase2}.
The phase shift extracted from the same fit using Eq.\eqref{phase2} is illustrated in Fig.~\ref{phaseshiftplot3D}, which has a much smaller uncertainty band compared to that in Fig.\ref{phaseshiftplot} for $3+1$ dimensions.

\newpage
\section{Summary and Conclusion}\label{sec:summary}

 In this work, we extend the relativistic ICF formalism developed in Refs.\cite{Guo:2023ecc,Guo:2024zal,Guo:2024pvt,Guo:2025ngh,Guo:2025vgk} to higher dimensions. 
 The central result is the ICF formula in Eq.\eqref{mainresult} or Eq.\eqref{mainresultEuclid}  that  relates the difference of interacting and non-interacting integrated correlation functions in finite volume to scattering phase shifts in infinite volume  through a weighted integral. 
 It can be regarded as a complementary method to the well-known L\"{u}scher formalism.
 The extension is an important step in  applying the formalism toward more realistic applications. 
 
 Specifically, we carry out Monte Carlo simulations of a complex $\phi^4$ field theory model on $L^3\times T$ lattices  in $3+1$ dimensions. Such a model simulates a relativistic scalar field interacting with a contact interaction potential.
 The main result is summarized in Fig.~\ref{diffCtplot} obtained on lattices up to $L=28$ and statistics on the order of 200k. 
 Despite the simplicity of the model, a decent amount of  computational resources is needed for the numerical demonstration.
 The convergence to the infinite volume limit is clearly visible but not ideal on large lattices. Nonetheless, the lattice data can be fitted against perturbation theory to extract the strength of contact interaction potential from which the scattering phase shift is determined.

Moreover, the features observed in $3+1$ dimensions are buttressed by a companion simulation in  $2+1$ dimensions where the computational burden is much alleviated and high statistics can be employed. The corresponding result in Fig.~\ref{diffCtplot3D} confirms the general conclusion in $3+1$ dimensions but the signal is much cleaner, and the phase shift can be extracted with a much smaller uncertainty.

These results inspire confidence that extracting phase shift in finite volume from the ICF approach is practical for quantum field theories in $3+1$ dimensions.
Reflecting on the experience, the accurate modeling of non-interacting two-particle correlation function in Eq.\eqref{mainresultEuclid} is a key ingredient for success. In $\phi^4$ theory, the non-interacting two-particle correlation function in Eq.(\ref{ct0A1analytic}) is an analytic function that depends only on the effective mass of the single particle; no other renormalization factors are required. The ultimate goal is to apply the formalism to simulate multi-hadron systems in lattice QCD. There, the hadron states emerge from dynamics based on quark and gluon degrees of freedom.  Hence the non-interacting two-particle correlation function depends on extra renormalization factors.  How the renormalization should be dealt with in real lattice QCD simulations is not clear at this stage and requires further studies and developments.

%\acknowledgments
%This research is supported by the U.S. National Science Foundation under grant PHY-2418937 (P.G.) and in part PHY-1748958 (P.G.), and the U.S. Department of Energy under grant  DE-FG02-95ER40907 (F.L. and A.A.). Numerical simulations are carried out on Perlmutter at NERSC and local clusters at DSU and GWU.

\acknowledgments
This research is supported by the U.S. National Science Foundation under grant PHY-2418937 (P.G.) and in part Grant No. PHY-2309135 to the Kavli Institute for Theoretical Physics (KITP) (P.G.), and the U.S. Department of Energy under grant  DE-FG02-95ER40907 (F.L. and A.A.). Numerical simulations are carried out on Perlmutter at NERSC and local clusters at DSU and GWU.

\appendix

\section{Relativistic  scattering solutions with a contact interaction in infinite volume}\label{scattsolutionsinfvol}

\subsection{Relativistic Lippmann-Schwinger like equation and scattering solutions}
For the complex $\phi^4$ theory   with an  instantaneous and contact interaction in the form of Eq.\eqref{contact}, the scattering dynamics can be    well described by relativistic Lippmann-Schwinger (LS) like equation, see e.g. Refs.~\cite{Guo:2019ogp,Guo:2020kph}. The relativistic wave function with an incoming plane wave, $e^{i \mathbf{ k} \cdot \mathbf{ r}}$, satisfies 
\begin{equation}
\psi^{(\infty)}_{E_k}  (\mathbf{ r}, \mathbf{ k})=  e^{i \mathbf{ k} \cdot \mathbf{ r}} +  G^{(0, \infty)}(\mathbf{ r}, E_k) V_0 \psi^{(\infty)}_{E_k}  (\mathbf{ 0}, \mathbf{ k}),
\end{equation}
where  $\mathbf{ r}$ and $ \mathbf{ k}$ are the relative coordinate and momentum of two particles in center of mass (CM) frame,  and $V_0$ denotes the bare contact interaction strength.  The total energy of two identical scalar particles with mass of $m$ is related to momentum $\mathbf{ k}$ via the relativistic dispersion relation,
\begin{equation}
E_k = 2 \sqrt{\mathbf{ k}^2 + m^2} .
\end{equation}
   The free-particle Green's function is defined by,
\begin{align}
G^{(0, \infty)} (\mathbf{ r}; E)  =  \int_{-\infty}^\infty \frac{d \mathbf{ q}}{(2\pi)^3}  \frac{1}{ \sqrt{\mathbf{ q}^2+ m^2}} \frac{e^{i \mathbf{ q} \cdot  \mathbf{ r}} }{E^2 -  E_q^2 + i 0 }. \label{infinitevolumefreeGreen}
\end{align}

With a contact interaction, only $S$-wave will contribute to non-trivial scattering solutions. By introducing the $S$-wave scattering amplitude, 
\begin{equation}
t(E) = - \frac{1}{\frac{1}{V_0} - G^{(0, \infty)} (\mathbf{ 0}; E) }, \label{scattamp}
\end{equation}
  the scattering solution for  wave function  is  given by
\begin{equation}
\psi^{(\infty)}_{E_k}  (\mathbf{ r}, \mathbf{ k})=  e^{i \mathbf{ k} \cdot \mathbf{ r}} -  t(E_k) G^{(0, \infty)} (\mathbf{ r};  E_k)  .
\end{equation}
The Green's  function  $G^{(0, \infty)} (\mathbf{ 0}; E)$  is an analytic function of $E^2$ with a branch cut starting at threshold $4m^2$,
   \begin{equation} 
G^{(0, \infty)} (\mathbf{ 0}; E)   =  \frac{ 1 }{16 \pi^2} \int_{4 m^2}^\infty d s'      \frac{  \sqrt{1 -  \frac{ 4 m^2 }{s'} }   }{E^2 -  s' + i 0 }.
\end{equation}
The imaginary part of Green's function at $\mathbf{ r} = \mathbf{ 0}$ is finite,
   \begin{equation} 
Im \left [ G^{(0, \infty)} (\mathbf{ 0}; E)  \right ]  =  \begin{cases}  -  \frac{ 1 }{16 \pi }      \sqrt{1 -  \frac{ 4 m^2 }{E^2} } , & \text{if} \ \ E > 4 m^2; \\ 0, & \text{otherwise}. \end{cases}
\end{equation}
However, the real part of Green's function at $\mathbf{ r} = \mathbf{ 0}$ diverges logarithmically and is cutoff dependent, see Eq.(B4) in Ref.\cite{Guo:2020kph}, 
\begin{equation}
 Re \left [  G^{(0, \infty)} (\mathbf{ 0}; E)       \right ]    =  Re \left [ G(E) \right ] - \frac{1}{8\pi^2} \ln \frac{2 \Lambda}{m} ,  
\end{equation}
where $\Lambda$ denotes the momentum cutoff.   The function   $G(E)$ in this equation is   divergence-free and well-defined,
\begin{equation}
  G(E)  =    \frac{\rho(E)}{ \pi }  \ln \frac{\rho(E) +1}{ \rho(E) -1}   ,
\end{equation}
where
  \begin{equation} 
\rho(E) =   \frac{ 1 }{16 \pi }      \sqrt{1 -  \frac{ 4 m^2 }{E^2} } . \label{infphasefactor}
\end{equation}
The divergence term of  Green's function  in the definition of scattering amplitude can be absorbed by redefining the bare coupling strength,
\begin{equation}
\frac{1}{V_0} = \frac{1}{V_R(\mu)} + G(\mu) - \frac{1}{8\pi^2} \ln \frac{2 \Lambda}{m} ,
\end{equation}
where $V_R(\mu)$ is the renormalized coupling strength at a renormalization scale $\mu $ with $ \mu $ chosen below two-particle threshold. Hence we find
\begin{equation}
t(E)   = - \frac{1}{\frac{1}{V_R(\mu)}  - G(E) + G(\mu) } .
\label{t1}
\end{equation}

The scattering amplitude is typically parameterized by a phase shift,
\begin{equation}
t(E)  = \frac{1}{\rho(E)} \frac{e^{2 i \delta(E)} -1}{2i} = \frac{1}{\rho(E)} \frac{1}{\cot \delta(E) - i }, \label{scattampparametrization}
\end{equation}
so the $S$-wave phase shift is given by
\begin{equation}
\delta (E) = \cot^{-1}  \left [ - \frac{\frac{1}{V_R(\mu)}  -  Re \left [ G(E) \right ]+ G(\mu) }{\rho(E)} \right ] .
\end{equation}

 The  Muskhelishvili-Omn\`es (MO) representation  \cite{muskhelishvili1941application,Omnes:1958hv,Guo:2022row}  of scattering amplitude is associated with the phase shift  via a dispersion integral relation by
\begin{align}
t(E)   = N e^{ \frac{1}{\pi} \int_{4 m^2}^\infty d s \frac{ \delta(\sqrt{s}) }{ s - E^2 - i 0} } , \label{MOscattamp}
\end{align}
where 
\begin{align}
N= t( \mu )e^{ - \frac{1}{\pi} \int_{4 m^2}^\infty d s \frac{ \delta(\sqrt{s}) }{ s -  \mu^2  } }, 
\end{align}
is a constant normalization factor.

Using analytic expression of the scattering amplitude, we can verify that  the scattering amplitude satisfies a generalized unitarity relation,
\begin{align}
& t(E) - t^*(E') \nonumber \\
& = \left [G^{(0, \infty)} ( \mathbf{ 0}; E)  - G^{(0, \infty) *} ( \mathbf{ 0}; E')  \right ]t^*(E') t(E). \label{unitarityrelation}
\end{align} 
  The ultraviolet divergence of Green's function at $\mathbf{ r} = \mathbf{ 0}$ is cancelled out on the right-hand side of Eq.(\ref{unitarityrelation}), and the generalized unitarity relation is free of divergence.
  This relation guarantees that the scattering amplitude in Eq.\eqref{t1} and Eq.\eqref{scattampparametrization} is well defined.

\subsection{Two-particle Green's function solution with a contact interaction}

The solution of relativistic two-particle Green's function with a contact interaction  is given by Dyson equation,
\begin{align}
& G^{(\infty)} (\mathbf{ r},\mathbf{ r'} ; E)  \\
& =  G^{(0, \infty)} (\mathbf{ r}-\mathbf{ r'}; E)  + G^{(0, \infty)} (\mathbf{ r}; E)  V_0 G^{(\infty)} (\mathbf{ 0},\mathbf{ r'} ; E).  \nonumber
\end{align}
Using Eq.\eqref{scattamp}, it can be written in terms of $t(E)$  matrix as, 
\begin{align}
& G^{(\infty)} (\mathbf{ r},\mathbf{ r'} ; E)  -  G^{(0, \infty)} (\mathbf{ r}-\mathbf{ r'}; E)  \nonumber \\
&  =  - G^{(0, \infty)} (\mathbf{ r}; E) t(E)   G^{(0, \infty)} (\mathbf{ r'} ; E). \label{infinitevolumeGreendiff}
\end{align}
The spectral representation of Green's function is given by,
\begin{align}
& G^{(\infty)} (\mathbf{ r}, \mathbf{ r'} ; E) \label{green2r} \\
& =  \int_{-\infty}^\infty \frac{d \mathbf{ q}}{(2\pi)^3}  \frac{1}{ \sqrt{\mathbf{ q}^2+ m^2}} \frac{ \psi^{(\infty)}_{E_q}  (\mathbf{ r}, \mathbf{ q}) \psi^{(\infty)*}_{E_q}  (\mathbf{ r'}, \mathbf{ q})   }{E^2 -  E_q^2 + i 0 },  \nonumber
\end{align}
where $E_q = 2 \sqrt{\mathbf{ q}^2+ m^2}$.

\subsection{Relativistic integrated Green's function and generalized Krein’s theorem}\label{relatFriedelformula}

Fourier transforming   Eq.(\ref{infinitevolumeGreendiff}) into momentum space, we find,
\begin{align}
& \sqrt{p^2+ m^2} \left [ \widetilde{G}^{(\infty)} (\mathbf{ p},\mathbf{ p}  ; E) - \widetilde{G}^{(0, \infty)} (\mathbf{ p}-\mathbf{ p}; E)  \right ]  \nonumber \\
&  =      t(E)    \frac{d}{d E^2} \left [  \frac{1}{ \sqrt{p^2+ m^2}} \frac{1 }{E^2 - E_p^2  + i 0 }   \right ]   , \label{green2p}
\end{align}
where the  Green's function in momentum space is defined by,
\begin{equation}
\widetilde{G}^{(\infty)} (\mathbf{ p},\mathbf{ p'} ; E)  = \int_{-\infty}^\infty d \mathbf{ r} d \mathbf{ r'}  e^{i \mathbf{ p} \cdot \mathbf{ r}} G^{(\infty)} (\mathbf{ r}, \mathbf{ r'} ; E)  e^{- i \mathbf{ p'} \cdot  \mathbf{ r'} }.
\end{equation}
The difference of  integrated Green's functions is therefore given   by 
\begin{align}
& \int_{-\infty}^\infty \frac{d \mathbf{ p}}{(2\pi)^3} \sqrt{p^2+ m^2}   \left [ \widetilde{G}^{(\infty)} (\mathbf{ p},\mathbf{ p}  ; E) - \widetilde{G}^{(0, \infty)} (\mathbf{ 0}; E)  \right ]  \nonumber \\
&  =    t(E)    \frac{d}{d E^2}   G^{(0, \infty)}(\mathbf{ 0}, E)   =   -  \frac{d}{d E^2}  \ln \left [   t(E)  \right ] .
\end{align}
Using   Eq.(\ref{MOscattamp}), the integrated Green's function is associated with the scattering phase shift by
\begin{align}
& \int_{-\infty}^\infty \frac{d \mathbf{ p}}{(2\pi)^3}  \sqrt{p^2+ m^2}  \left [ \widetilde{G}^{(\infty)} (\mathbf{ p},\mathbf{ p}  ; E) - \widetilde{G}^{(0, \infty)} (\mathbf{ 0} ; E)  \right ]  \nonumber \\
&  =    -   \frac{1}{\pi} \int_{4 m^2}^\infty d s \frac{ \delta(\sqrt{s}) }{ (s - E^2 - i 0)^2 }   , \label{Kreintheorem}
\end{align}
and we also have a relation for the imaginary part,
\begin{align}
&  \int_{-\infty}^\infty \frac{d \mathbf{ p}}{(2\pi)^3}  \sqrt{p^2+ m^2}  Im \left [ \widetilde{G}^{(\infty)} (\mathbf{ p},\mathbf{ p}  ; E) - \widetilde{G}^{(0, \infty)} (\mathbf{ 0} ; E)  \right ]  \nonumber \\
&  =    -   \frac{ d \delta(E) }{d E^2} . \label{Friedelformula}
\end{align}
The relations in Eq.(\ref{Friedelformula})  and Eq.(\ref{Kreintheorem}) are a relativistic generalization of Friedel formula \cite{Friedel1958} and Krein’s theorem \cite{zbMATH03313022,krein1953trace} respectively, also see discussions in Ref.~\cite{Guo:2022row}.  
Using the partial-wave expansion of Green's function,
 \begin{equation}
 \widetilde{G}^{(\infty)} (\mathbf{ p}, \mathbf{ p'} ; E) = \sum_{l m_l} Y_{l m_l} (\mathbf{ \hat{p}})    \widetilde{G}_l^{(\infty)} ( p ,  p' ; E)  Y^*_{l m_l} (\mathbf{ \hat{p'}})    ,
\end{equation}
the relation can be written in terms of partial waves,
\begin{align}
&  \int_{0}^\infty \frac{p^2 d  p }{(2\pi)^3}  \sqrt{p^2+ m^2}  \left [ \widetilde{G}_l^{(\infty)} (p, p ; E) - \widetilde{G}_l^{(0, \infty)} ( 0; E)  \right ]  \nonumber \\
&  =    -   \frac{1}{\pi} \int_{4 m^2}^\infty d s \frac{ \delta_l (\sqrt{s}) }{ (s - E^2 - i 0)^2 }   .
\end{align}

\section{The integrated correlation function formalism for a complex $\phi^4$ lattice  model}\label{modelsetup}  
In this section, we present some technical details of the integrated correlation function formalism for a simple relativistic lattice model in $3+1$ spacetime dimensions. Its $1+1$ correspondence can be found in Ref.~\cite{Guo:2024zal}.
We consider the complex $\phi^4$ model with interaction of charged scalar particles via a short range potential whose classical action in four-dimensional Minkowski space-time is,  
\begin{align}
S &=  \frac{1}{2} \int_{-\infty}^\infty d t \int_{0}^L d \mathbf{ x}  \left [  \frac{\partial \phi^* }{\partial t} \frac{\partial \phi }{\partial t} -     \nabla \phi^*  \cdot \nabla    \phi   -  m_0^2 | \phi |^2 \right ] \nonumber \\
& - \frac{g_0}{4 !}   \int_{-\infty}^\infty d t \int_{0}^L d \mathbf{ x}   |  \phi (\mathbf{ x}, t)   |^4,   \label{Minkowskiaction}
\end{align}
where $m_0$ and $g_0$ are the bare mass and coupling strength of charged scalar field. The action is written down in a finite box of size $L$ and continuous time $t$. The complex     $\phi (\mathbf{ x},t)$  field operator    satisfies the periodic boundary condition
\begin{equation}
\phi (\mathbf{ x} + L \hat{\mathbf{ x}},t) = \phi (\mathbf{ x},t) ,
\end{equation}
where  $\hat{\mathbf{ x}}$ stands for the spatial dimensional unit vector.
Its  discretization in Euclidean spactime in Eq.\eqref{Euclideanaction} is used in Monte Carlo  simulations.

\subsection{Two-particle correlation function and its spectral representation}
The two-particle interaction is encoded in the time dependence of correlation functions defined by,
\begin{align}
C^{(2\phi)} (\mathbf{ r}t; \mathbf{ r'}0) & = \theta (t) \langle 0 |   \mathcal{O} (\mathbf{ r}, t )     \mathcal{O}^\dag (\mathbf{ r'} ,0)   | 0 \rangle \nonumber \\
& + \theta (-t) \langle 0 |    \mathcal{O}^\dag (\mathbf{ r'} ,0)    \mathcal{O} (\mathbf{ r}, t )   | 0 \rangle, 
\end{align}
 where  $    \mathcal{O}^\dag (\mathbf{ r'} ,0)  $  and $ \mathcal{O} (\mathbf{ r}, t ) $    creates and annihilates   two identical charged particles with relative coordinates and time  at $(\mathbf{ r'}, 0)$ and  $(\mathbf{ r}, t ) $, respectively.   The creation operator   after projecting out center of mass motion (CM) in the rest frame    is defined by,
 \begin{equation}
    \mathcal{O}^\dag (\mathbf{ r} ,t) =  \frac{1}{\sqrt{2}} \int_0^L \frac{d \mathbf{ x}_2 }{\sqrt{L^3}} \phi(\mathbf{ r}+\mathbf{ x}_2,t) \phi(\mathbf{ x}_2,t). \label{reltwoparticleoperator}
 \end{equation}
The factor $1/\sqrt{2}$ takes into account the exchange symmetry of two distinguishable charged particles. Inserting a complete energy basis, $\sum_n | E_n \rangle \langle E_n | = 1$,  and also defining  two-particle and two-antiparticle relative wave functions,
\begin{align}
& \langle E_n  |   \mathcal{O}^\dag (\mathbf{ r'} ,0) | 0\rangle   = \frac{1}{\sqrt{L^3}} \frac{ \psi^{(L)*}_{E_n} (\mathbf{ r'})}{E_n}  ,    \nonumber \\
& \langle E_n  |   \mathcal{O}  (\mathbf{ r} ,t) | 0\rangle    = \frac{1}{\sqrt{L^3}} \frac{ \psi^{(L)*}_{E_n} (\mathbf{ r})}{E_n} e^{ i  E_n t} ,
\end{align}
 where we have assumed that two-particle and two-antiparticle wave functions are identical,
 the spectral representation of relativistic two-particle correlation function can be written as,
\begin{align}
C^{(2\phi)}(\mathbf{ r}t; \mathbf{ r'}0) & = \frac{\theta (t)}{L^3} \sum_n \frac{ \psi^{(L)}_{E_n} (\mathbf{ r})}{E_n} \frac{ \psi^{(L)*}_{E_n} (\mathbf{ r'})}{E_n} e^{- i  E_n t} \nonumber \\
& +  \frac{\theta (-t)}{L^3} \sum_n  \frac{ \psi^{(L)}_{E_n} (\mathbf{ r'})}{E_n}  \frac{ \psi^{(L) *}_{E_n} (\mathbf{ r})}{E_n}e^{ i  E_n t}. \label{twocorrspectral}
\end{align}
In the absence of interactions, the  correlation function takes the form of,
\begin{align}
& C^{(0,2\phi)} (\mathbf{ r}t; \mathbf{ r'}0)  \nonumber \\
&  = \frac{1}{L^3} \sum_{\mathbf{ p} = \frac{2\pi \mathbf{ n}}{L}, \mathbf{ n} \in \mathbb{Z}^3} \frac{  \cos ( \mathbf{ p} \cdot  \mathbf{ r} ) }{ E^{(0)}_p } \frac{  \cos (\mathbf{ p} \cdot  \mathbf{ r'}) }{E^{(0)}_p } e^{- i  E^{(0)}_p | t| },
\end{align}
where free two-particle energies are $E^{(0)}_p  =2  \sqrt{\mathbf{ p}^2 + m^2}$  with $\mathbf{ p} = \frac{2\pi \mathbf{ n}}{L}, \mathbf{ n} \in \mathbb{Z}^3$.

The  two-particle relative wave function  satisfies relativistic LS-like equation, see e.g.  Appendix \ref{scattsolutionsinfvol},
\begin{equation}
 \psi^{(L)}_{E} (\mathbf{ r}) =   V_0 G^{(0, L)} (\mathbf{ r}  ; E)     \psi^{(L)}_{E} (\mathbf{ 0}). \label{finitevolumeLSeq}
\end{equation}
The relativistic finite volume Green's function is defined by, see e.g. Refs.~\cite{Guo:2019ogp,Guo:2020kph},
\begin{equation}
G^{(0, L)} (\mathbf{ r}; E) = \frac{1}{L^3} \sum_{\mathbf{ q} = \frac{2\pi \mathbf{ n}}{L}, \mathbf{ n} \in \mathbb{Z}^3 } \frac{  1}{ \omega_q   } \frac{e^{i \mathbf{ q} \cdot  \mathbf{ r}}}{E^2 - ( 2 \omega_q )^2}, \label{finitevolumefreeGreen}
\end{equation}
where  $ \omega_q = \sqrt{\mathbf{ q}^2 + m^2}$  is the energy of a single particle with momentum  $\mathbf{ q}$. 
The two-particle wave function must be symmetric     due to Bose symmetry,
\begin{equation}
 \psi^{(L)}_{E_n} (-\mathbf{ r}) =  \psi^{(L)}_{E_n} (\mathbf{ r}).
\end{equation}
The relativistic wave function is normalized in momentum space by
\begin{equation}
\frac{1}{L^3} \sum_{\mathbf{ p} = \frac{2\pi \mathbf{ n}}{L}, \mathbf{ n} \in \mathbb{Z}^3 } \frac{1}{2 \omega_p} \widetilde{\psi}^{(L)}_{E_n} (\mathbf{ p})  \widetilde{\psi}^{(L)*}_{E_{n'}} (\mathbf{ p}) = E_n L^3 \frac{ \delta_{n, n'}  +  \delta_{n, -n'}}{2}, \label{normalizationwavfunc}
\end{equation}
where the Fourier transform of wave function is defined by
\begin{equation}
 \frac{  \widetilde{\psi}^{(L)}_{E_n} (\mathbf{ p}) }{2 \omega_p}= \int_0^L d \mathbf{ r}  \psi^{(L)}_{E_n} (\mathbf{ r}) e^{i \mathbf{ p} \cdot \mathbf{ r}}.
\end{equation}

The normalization relation of wave function in Eq.(\ref{normalizationwavfunc}) leads to cubic symmetry $A_1^+$ irrep projected integrated correlation function,
\begin{equation}
C_{A_1^+}^{(2\phi)}(t) =    \frac{1}{L^3}  \sum_{\mathbf{ p} = \frac{2\pi \mathbf{ n}}{L}, \mathbf{ n} \in \mathbb{Z}^3} 2 \omega_p  \widetilde{C}_{A_1^+}^{(2\phi)}(\mathbf{ p}t; \mathbf{ p}0) =  \sum_{n } \frac{ e^{- i  E_n | t | } }{E_n} , \label{integratedCt}
\end{equation}
where $\widetilde{C}_{A_1^+}^{(2\phi)}$  is the integrated correlation function in momentum space.   Its Fourier transform and $A_1^+$ irrep projection are performed by,
\begin{equation}
 \widetilde{C}^{(2\phi)}(\mathbf{ p}t; \mathbf{ p'}0) =  \int_0^L d \mathbf{ r} d \mathbf{ r'}  e^{i \mathbf{ p} \cdot  \mathbf{ r}}   C^{(2\phi)}(\mathbf{ r}t; \mathbf{ r'}0)  e^{- i \mathbf{ p'} \cdot  \mathbf{ r'}}, \label{FourierTFdef}
\end{equation}
and
\begin{equation}
 \widetilde{C}_{A_1^+}^{(2\phi)}(\mathbf{ p}t; \mathbf{ p'}0) = \frac{1}{48^2} \sum_{(g, g') \in \mathcal{G}}   \widetilde{C}^{(2\phi)}(g \mathbf{ p}t;  g' \mathbf{ p'}0).
\end{equation}
The $(\mathcal{G}, g)$ stand for   the cubic symmetry  octahedral group $O_h$ and its elements, see e.g. Ref.~\cite{Doring:2018xxx,Cornwell:1997ke,Guo:2020kph}.  We remark that only $A_1^+$ irrep is projected in this work since  the $\phi^4$ theory is equivalent to a contact interaction theory, hence only $S$-wave contributes to the scattering dynamics.

Using the relation,
\begin{align}
&  \left  | \frac{1}{48} \sum_{g\in \mathcal{G}} \left [ \int d \mathbf{ r} e^{i  g \mathbf{ p}_0 \cdot \mathbf{ r}} \cos (\mathbf{ k} \cdot \mathbf{ r}) \right ] \right   |^2 \nonumber \\
&  = \begin{cases}  \frac{L^6}{\vartheta^2 (\mathbf{ p}_0)}, & \text{if} \ \mathbf{ k} \in g \mathbf{ p}_0 , \ \text{where} \  g \in \mathcal{G} , \\  0, & \text{otherwise}, \end{cases}
\end{align}
where $\mathbf{ p}_0$ denote a single reference vector,   the set of momenta represented by $\mathbf{ p}_0$ via $\mathbf{ p} = g \mathbf{ p}_0 (g \in \mathcal{G})$ is called the ``star" of $\mathbf{ p}_0$,  and   $\vartheta(\mathbf{ p}_0)$ is the multiplicity of distinct momenta within the set represented by $\mathbf{ p}_0$, see e.g. Ref.~\cite{Doring:2018xxx,Cornwell:1997ke,Guo:2020kph},  we can thus show that
\begin{equation}
C_{A_1^+}^{(0, 2\phi)}(t)   =  \sum_{\mathbf{ p}_0 } \frac{ e^{- i   E^{(0)}_{\mathbf{ p}_0} | t | } }{ E^{(0)}_{\mathbf{ p}_0} } . \label{integratedC0t}
\end{equation} 
The reference  vectors, $\mathbf{ p}_0$'s, may be chosen as
\begin{equation}
\mathbf{ p}_0  = \frac{2\pi}{L} (i, j , k),  \ \ \ \ ( i, j, k ) \in [0, 1, \cdots] \, \&  \, (i \leqslant j \leqslant k).
\end{equation}
The sum of momenta over a general function can be reorganized as follows:
\begin{equation}
\sum_{\mathbf{ p} = \frac{2\pi \mathbf{ n}}{L}, \mathbf{ n} \in \mathbb{Z}^3} f(\mathbf{ p}) =  \sum_{\mathbf{ p}_0 }  \vartheta(\mathbf{ p}_0) \frac{1}{48} \sum_{g \in \mathcal{G}} f(g \mathbf{ p}_0) .
\end{equation}
Hence instead of using label $n$, we can use $\mathbf{ p}_0$ to label all the eigenenergies: $ n \rightarrow \mathbf{ p}_0$ and $E_n \rightarrow E_{\mathbf{ p}_0}$.  We thus should also have a similar relation as in Eq.(\ref{integratedC0t}) for interacting wave function, 
\begin{align}
&  \left  | \frac{1}{48} \sum_{g\in \mathcal{G}} \left [ \int d \mathbf{ r} e^{i  g \mathbf{ p}_0 \cdot \mathbf{ r}}    \psi^{(L)}_{E_{\mathbf{ k}}} (\mathbf{ r})  \right ] \right   |^2 \nonumber \\
&  = \begin{cases}  \frac{L^6}{\vartheta^2 (\mathbf{ p}_0)}, & \text{if} \ \mathbf{ k} \in g \mathbf{ p}_0 , \ \text{where} \  g \in \mathcal{G} , \\  0, & \text{otherwise}. \end{cases}
\end{align}
The spectral representation of difference of interacting and non-interacting integrated correlation functions in $A_1^+$ irrep is thus given by
\begin{align}
& \triangle C_{A_1^+}^{(2\phi)}(t)   \nonumber \\
 &= \frac{1}{L^3} \sum_{\mathbf{ p} = \frac{2\pi \mathbf{ n}}{L}, \mathbf{ n} \in \mathbb{Z}^3 }   2 \omega_p  \left [   \widetilde{C}_{A_1^+}(\mathbf{ p} t; \mathbf{ p} 0) -   \widetilde{C}_{A_1^+}^{(0)}(\mathbf{ p} t; \mathbf{ p} 0)  \right ] \nonumber \\
& =   \sum_{\mathbf{ p}_0 }  \left [ \frac{ e^{- i  E_{\mathbf{ p}_0} t} }{E_{\mathbf{ p}_0}}  -  \frac{ e^{- i  E^{(0)}_{\mathbf{ p}_0} t} }{E^{(0)}_{\mathbf{ p}_0}}  \right ] .  \label{diffintegratedcorrfuncdef}
\end{align}
Therefore, there are two ways to obtain the difference of integrated correlations in the main relation in Eq.\eqref{mainresultEuclid}, 
either by direct calculation plus Fourier transform, or by energy spectrum if it is available.

\subsection{Relating integrated correlation functions to scattering phase shift}

Conisder the momentum-space spectral representation of   two-particle correlation function along the diagonal,
\begin{align}
&  \widetilde{C}^{(2\phi)}(\mathbf{ p} t; \mathbf{ p}0)  \nonumber \\
&= \frac{1}{L^3} \sum_n   \frac{1}{E_n^2  }  \frac{ |  \widetilde{\psi}^{(L)}_{E_n} (\mathbf{ p}) |^2}{E^2_p}     \left [\theta (t)  e^{- i  E_n t} + \theta(-t)  e^{ i  E_n t} \right ].
\end{align}
Using the identity,
\begin{equation}
i \int_{-\infty}^{\infty} \frac{d \lambda}{2\pi} \frac{e^{- i \lambda t}}{\lambda + i  0 } = \theta (t),  
\end{equation}
the diagonal element of   two-particle correlation function  can be related to two-particle Green's function by,
\begin{equation}
\widetilde{C}^{(2\phi)} (\mathbf{ p} t; \mathbf{ p}0)  =    i \int_{-\infty}^{\infty} \frac{d \lambda}{2\pi}  \widetilde{G}^{(L)} (\mathbf{ p} ,  \mathbf{ p}; \lambda)  e^{- i \lambda t}    .
\end{equation}
The spectral representation of two-particle Green's function in finite volume is given by
\begin{equation}
G^{(L)}(\mathbf{ r}, \mathbf{ r'}; E)   =      \frac{1}{L^3} \sum_{\mathbf{ q} = \frac{2\pi \mathbf{ n}}{L}, \mathbf{ n}\in \mathbb{Z}^3 } \frac{1   }{\omega_q}   \frac{    \psi^{(L)}_{E_q} (\mathbf{ r})   \psi^{(L)*}_{E_q} (\mathbf{ r'}) }{  E^2 -  E_q^2  + i 0 }   ,
\end{equation}
and $E_q = 2 \omega_q = 2 \sqrt{q^2+m^2}$. 
It is the lattice version of Eq.\eqref{green2r}.
The momentum space two-particle Green's function, $\widetilde{G}^{(L)}(\mathbf{ p}, \mathbf{ p'}; E) $, is the Fourier transform of coordinate space two-particle Green's function,
\begin{equation}
\widetilde{G}^{(L)}(\mathbf{ p}, \mathbf{ p'}; E)   =      \int_0^L d \mathbf{ r}  d \mathbf{ r'} e^{i \mathbf{ p} \cdot \mathbf{ r}} G^{(L)}(\mathbf{ r}, \mathbf{r'}; E)  e^{- i \mathbf{ p'} \cdot  \mathbf{ r'}} ,
\end{equation}
which is lattice version of Eq.\eqref{green2p}.
The  difference of integrated correlation functions in $A_1^+$ irrep is  related to finite volume Green's functions by
\begin{align}
 & \triangle C_{A_1^+}^{(2\phi)} (t)  = i \int_{-\infty}^{\infty} \frac{d \lambda}{2\pi}   e^{- i \lambda t}   \\
&  \times   \frac{1}{L^3} \sum_{\mathbf{ p} = \frac{2\pi \mathbf{ n}}{L}, \mathbf{ n} \in \mathbb{Z}^3}   2 \omega_p \left [\widetilde{G}_{A_1^+}^{(L)}(\mathbf{ p}, \mathbf{ p}; \lambda)-\widetilde{G}_{A_1^+}^{(0, L)}(\mathbf{ 0} ; \lambda)  \right ] . \nonumber 
\end{align}
 The $A_1^+$ irrep projected  Green's functions are defined by
 \begin{equation}
 \widetilde{G}_{A_1^+}^{(L)}(\mathbf{ p}, \mathbf{ p'}; \lambda) = \frac{1}{48^2} \sum_{ (g, g' ) \in \mathcal{G}} \widetilde{G}^{(L)}( g \mathbf{ p}, g' \mathbf{ p'}; \lambda).
 \end{equation}

Using relativistic Friedel formula and Krein's theorem relations in infinite volume, see Appendix \ref{relatFriedelformula},
    \begin{align} 
  &  \int_{-\infty}^\infty   \frac{d \mathbf{ p}}{(2\pi)^3}     \omega_p   \left [  \widetilde{G}^{(\infty)} (\mathbf{ p} , \mathbf{ p}; E)    -  \widetilde{G}^{(0, \infty)} (\mathbf{ 0}  ; E)  \right ]  \nonumber \\
  &  = -       \frac{1}{\pi} \int_{4 m^2}^\infty d s \frac{ \delta(\sqrt{s}) }{  (s - E^2 - i 0 )^2}    ,
\end{align}
where $\delta (E)$ is the $S$-wave scattering phase shift of two scalar particles,  the difference of integrated correlation functions thus approaches its infinite volume limit by
\begin{align}
&  \triangle C_{A_1^+}^{(2\phi)} (t) \\
 & \stackrel{L \rightarrow \infty}{\rightarrow}     -    \frac{1}{\pi} \int_{4 m^2}^\infty d s \delta(\sqrt{s})  \left [    i \int_{-\infty}^{\infty} \frac{d \lambda}{\pi}     \frac{  e^{- i \lambda t} }{ (s - \lambda^2 - i 0)^2}  \right ]  .  \nonumber 
\end{align}
Completing the integration in the bracket,  we find a compact relation in Eq.\eqref{mainresult} that closely resembles its $1+1$ dimensional counterpart in Ref.~\cite{Guo:2024zal}.

\subsection{Quantization condition in finite volume}\label{exactmodelcontactpot}
 With a contact interaction defined in Eq.\eqref{contact}, the quantization condition can be obtained via   Eq.(\ref{finitevolumeLSeq}),  
\begin{equation}
 \frac{1}{V_0}  =   G^{(0, L)} ( \mathbf{ 0} ; E)  ,
\end{equation}
where the finite volume free two-particle Green's function is given in Eq.(\ref{finitevolumefreeGreen}). Using  Eq.(\ref{scattamp}) and Eq.(\ref{scattampparametrization}), the potential strength $V_0$ can be related to infinite volume free particles Green's function and scattering phase shift by,
\begin{equation}
 \frac{1}{V_0}   = Re \left [ G^{(0,\infty)} ( \mathbf{ 0} ; E)  \right ] - \rho(E) \cos \delta(E), \label{phaseshift}
\end{equation}
where  the analytic expression of $G^{(0,\infty)} ( 0 ; E)$ and $\rho(E)$ are given  in Eq.(\ref{infinitevolumefreeGreen}) and Eq.(\ref{infphasefactor}), respectively. The quantization condition can be rewritten in a compact form known as L\"uscher  formula,
\begin{equation}
  \cos \delta(E) =  \frac{Re \left [ G^{(0, \infty)} ( \mathbf{ 0} ; E)  \right ]  - G^{(0,L)} ( \mathbf{ 0} ; E) }{\rho(E)}, \label{lusherformula}
\end{equation}
the right hand side of Eq.(\ref{lusherformula}) is typically referred to zeta function, see e.g. Ref.\cite{Luscher:1990ux}. 
The above discussion demonstrates the indirect connection via Green's functions between the ICF formalism and the L\"{u}scher formalism.

\subsection{Perturbation result}\label{perturbationresult}

  The perturbation calculation of integrated correlation functions in a periodic box of size $L$  can be carried out in a similar fashion as in Ref.~\cite{Guo:2023ecc,Guo:2024zal}.  The leading order contribution of the difference of integrated correlation functions in finite volume is given by,
\begin{align}
& C^{(2\phi)} ( \mathbf{ r}, t; \mathbf{ r'}, 0 )  -  C^{(0, 2\phi)} ( \mathbf{ r}, t;\mathbf{  r' },0 )  \nonumber \\
&  \simeq  - \frac{ i V_R (0)}{L^3} \int_{0}^L  d \mathbf{ x}_2  \int_{0}^L  d \mathbf{ x'}_2  \int_{-\infty}^\infty d t' \int_0^L d \mathbf{ x''}   \nonumber \\
& \times    D_0^{-1} (\mathbf{ r}+ \mathbf{ x}_2 -\mathbf{ x''} , t - t''  ) D^{-1}_0 (\mathbf{ x}_2 -\mathbf{ x''} , t -t'' )   \nonumber \\
& \times    D_0^{-1} (\mathbf{ x''} - \mathbf{ r'}-  \mathbf{ x'}_2 ,  t''  ) D^{-1}_0 (\mathbf{ x''} -\mathbf{ x'}_2 ,   t'' ) , \label{perturbationcorrelation}
\end{align}
where $V_R(0)$ is the renormalized contact potential strength with renormalization scale chosen at $\mu =0$, and $D_0^{-1}$ the free two-particle propagator defined by,
\begin{equation}
D_0^{-1} (\mathbf{ x}   , t )   = i \int_{-\infty}^\infty  \frac{d  \epsilon }{2\pi} \frac{1}{L^3} \sum_{\mathbf{ k} = \frac{2\pi \mathbf{ n}}{L}, \mathbf{ n} \in \mathbb{Z}^3}  \frac{  e^{i k x  }  e^{i  \epsilon   t   }}{ \epsilon^2  -( k^2 +  m^2 )} . \label{invD0expression}
\end{equation}
After carrying out the space and time integrations, we find,
\begin{align}
& C^{(2\pi)} ( \mathbf{ r}, t; \mathbf{ r'}, 0 )  -  C^{(0, 2\phi)} ( \mathbf{ r}, t;\mathbf{  r' },0 )   \nonumber \\
&  \simeq -  i V_R(0)   \int_{-\infty}^\infty  \frac{d \epsilon }{2\pi}  e^{i \epsilon   t   } G^{(0, L)} (\mathbf{ r}; \epsilon ) G^{(0, L)} (\mathbf{ r'}; \epsilon)    ,
\end{align}
in terms of the finite volume free-particle Green's function defined in Eq.(\ref{finitevolumefreeGreen}). Using  the spectral representation in Eq.(\ref{diffintegratedcorrfuncdef}) and   carrying out the integration of $  \epsilon $,   we obtain the leading-order result from perturbation theory,  
\begin{equation}
 \triangle C_{pert}^{(2\phi)} (  t )      \simeq   -      \frac{V_R(0)}{L^3}  \sum_{\mathbf{ k} = \frac{2\pi \mathbf{ n}}{L}, \mathbf{ n} \in \mathbb{Z}^3}     \frac{   i t +  \frac{1}{E_k }   }{  E_k^3 }     e^{-  i E_k t}       ,  \label{perturbfinitevolume}
\end{equation}
where   $E_k =2 \sqrt{k^2+ m^2}$.  It approaches the infinite volume limit by,
 \begin{equation}
 \triangle C_{pert}^{(2\phi)} (  t )   \stackrel{L \rightarrow \infty}{\rightarrow}  -      V_R(0)  \int_{- \infty}^{\infty} \frac{d \mathbf{ k}}{(2\pi)^3}    \frac{  it  +  \frac{1}{E_k }   }{  E_k^3 }     e^{- i E_k t}        . \label{dCtfinitevolumpert}
\end{equation}

On the other hand,  using the perturbation expansion of scattering phase shift 
via Eq.\eqref{phaseshift} and Eq.\eqref{infphasefactor},
\begin{equation}
\delta(E_k )  \simeq -  V_R(0) \rho(E_k)     =  - \frac{V_R (0)}{16\pi} \frac{2 k}{E_k} ,
\end{equation}
we can verify that,
 \begin{align}
&  -   \frac{1}{\pi}   \int_{2m}^\infty   d  \epsilon  \delta (\epsilon)  \frac{d}{ d \epsilon}  \left (   \frac{   e^{-  it    \epsilon  t}  }{\epsilon}  \right ) \nonumber \\
& \simeq   - V_R(0)       \int_{-\infty}^\infty     \frac{     d \mathbf{ k} }{(2\pi)^3}     \left ( it + \frac{1}{E_k } \right )   \frac{   e^{-    i E_k  t}  }{E^3_k}    ,
\end{align}
which is Eq.(\ref{dCtfinitevolumpert}).

On the finite Euclidean spacetime lattice $L^3\times T$, the perturbation result for the difference of integrated correlation function takes the form,
\begin{equation}
\triangle C^{(2 \phi)}_{pert} (  \tau) =   \frac{1}{L^3} \sum_{\mathbf{ p} = \frac{2\pi \mathbf{ n}}{L}}^{\mathbf{ n} \in    [- \frac{L}{2} +1, \frac{L}{2}]^3 }  2 \omega^{(lat)}_\mathbf{ p}  \triangle \widetilde{C}_{pert}^{(2\phi)}(\mathbf{ p}, \tau)  , \label{dCpertfinitelatticespacing}
\end{equation}
where the momentum-space result before integration is,
\begin{equation}
\triangle \widetilde{C}^{(2 \phi)}_{pert} ( \mathbf{ p}, \tau) = - \frac{V_R(0)}{T}  \sum_{\omega = \frac{2\pi n}{L}}^{n \in [0, T-1]}  e^{i \omega \tau} \left [ G_{2\phi}  (\mathbf{ p}, \omega) \right ]^2.
\end{equation}
The finite volume two-particle Green's function is defined by,
\begin{align}
& G_{2\phi}  (\mathbf{ p}, \omega)  \nonumber \\
&= \frac{1}{T}  \sum_{\omega' = \frac{2\pi n'}{L}}^{n' \in [0, T-1]}   \frac{1}{ \sum_i (2  - 2   \cos   \mathbf{ k}_i) -2  \cos \omega'       + 2 \cosh m  } \nonumber \\
& \times  \frac{1}{ \sum_i ( 2  - 2   \cos   \mathbf{ k}_i ) -2  \cos (\omega - \omega' )      + 2 \cosh m  } .
\end{align}
In   the  limit of $T \rightarrow \infty$ and   zero  lattice spacing,  it takes the simple form,
\begin{equation}
  G_{2\phi}  (p, \omega)   \stackrel[a\rightarrow 0]{T \rightarrow \infty}{ \rightarrow }   \frac{1}{\omega_p} \frac{1}{\omega^2+ (2\omega_p)^2}  ,
\end{equation}
which leads to the difference of two-particle correlation functions,
\begin{equation}
\triangle \widetilde{C}^{(2 \phi)}_{pert} ( \mathbf{ p}, \tau)   \stackrel[a\rightarrow 0]{T \rightarrow \infty}{ \rightarrow }  - V_R(0)  \frac{\tau + \frac{1}{2\omega_p}}{(2\omega_p)^4} e^{- 2 \omega_p \tau} ,
\end{equation}
and its integrated version,
\begin{equation}
\triangle C^{(2 \phi)}_{pert} (  \tau) \stackrel[a\rightarrow 0]{T \rightarrow \infty}{ \rightarrow } -  \frac{V_R(0) }{L^3} \sum_{\mathbf{ p} = \frac{2\pi \mathbf{ n}}{L}}^{\mathbf{ n} \in    \mathbb{Z}^3 }    \frac{\tau + \frac{1}{2\omega_\mathbf{ p}}}{(2\omega_\mathbf{ p})^3} e^{- 2 \omega_\mathbf{ p} \tau}  .
\end{equation}
Hence, the perturbation result at the limit of zero lattice spacing in Eq.(\ref{perturbfinitevolume})  is recovered after analytic continuation $\tau \to i\,t$.

  \begin{figure}
\includegraphics[width=0.99\textwidth]{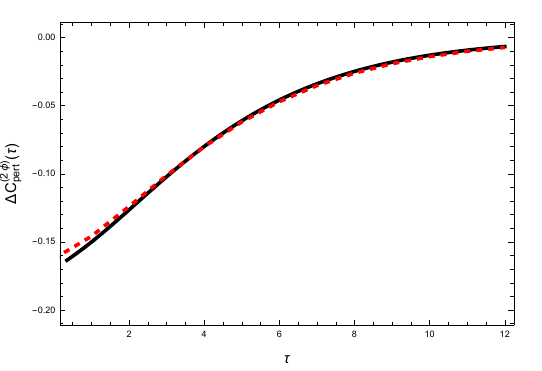}
 \caption{Numerical check of the spectral representation in Eq.\eqref{pert2}: left-hand-side (solid black) vs. right-hand-side (dashed red).  The parameters are:  $L=8$, $m=0.2$ and $V^{(L)}_R(0)=2$.  \label{sumdiffCtL8plot} }
 \end{figure}

  \begin{figure}
\includegraphics[width=0.99\textwidth]{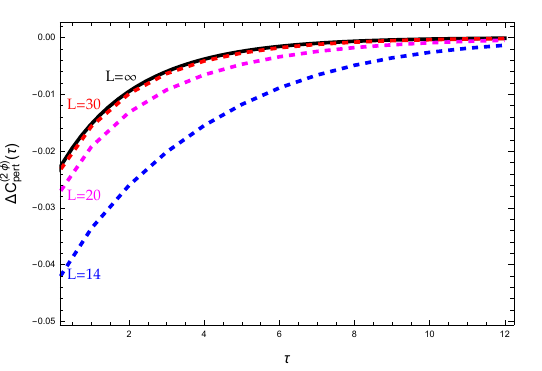}
 \caption{  Perturbation result on the right-hand-side of Eq.\eqref{pert2} for $L=14$ (dashed blue), $L=20$ (dashed magenta), and $L=30$ (dashed red) vs. infinite volume limit of $ \triangle C^{(2\phi)}( t) $ (solid black).  The other parameters are:  $m=0.2$ and $V^{(L)}_R(0)=2$.  \label{diffCtLpertplot} }
 \end{figure} 

\subsection{Numerics}\label{numericaltest}

As a short summary for this appendix, the  difference of integrated correlation functions in finite volume can be calculated by the spectral representation in Eq.\eqref{diffintegratedcorrfuncdef} and 
is related to phase shift  through a weighted integral, 
\begin{align}
   \triangle C_{A_1^+}^{(2\phi)}( t)  & \stackrel{t= - i \tau}{= }     \sum_{\mathbf{ p}_0  }    \left [    \frac{ e^{- E_{\mathbf{ p}_0}  \tau}  }{  E_{\mathbf{ p}_0}   } -     \frac{ e^{- 2 \omega_{\mathbf{ p}_0}   \tau}  }{  2 \omega_{\mathbf{ p}_0}  }    \right ]  \nonumber \\
   &   \stackrel[t=-i \tau]{L \rightarrow \infty}{\rightarrow}       \frac{1}{\pi} \int_{2 m}^\infty d \epsilon   \delta( \epsilon)     \left ( \tau + \frac{1}{\epsilon}     \right )  \frac{e^{ -   \epsilon \tau} }{ \epsilon} ,
\end{align}
where the eigenenergies $E_{\mathbf{ p}_0}$  are solutions of  L\"uscher  formula in Eq.(\ref{lusherformula}). For  the contact interaction potential, instead of using Eq.(\ref{lusherformula}), we will simply introduce a finite volume dependent  renormalized   coupling strength by  
\begin{equation}
 \frac{1}{V_0}  =  \frac{1}{V^{(L)}_R(0)} +  G^{(0, L)} ( \mathbf{ 0} ; 0)  ,
\end{equation}
  the renormalized quantization condition is thus given by
\begin{equation}
  \frac{1}{V^{(L)}_R(0)} = G^{(0, L)} ( \mathbf{ 0} ; E) -  G^{(0, L)} ( \mathbf{ 0} ; 0) .
\end{equation}
In Fig.~\ref{sumdiffCtL8plot}, we show that  the spectral representation of finite volume $ \triangle C_{A_1^+}^{(2\phi)}( t) $ matches well  with perturbation result:
\begin{align}
   &      \sum_{\mathbf{ p}_0  }    \left [    \frac{ e^{- E_{\mathbf{ p}_0}  \tau}  }{  E_{\mathbf{ p}_0}   } -     \frac{ e^{- 2 \omega_{\mathbf{ p}_0}   \tau}  }{  2 \omega_{\mathbf{ p}_0}  }    \right ]   \nonumber \\
   &   \stackrel{ V^{(L)}_R(0) \sim 0}{\rightarrow}        -      \frac{V^{(L)}_R(0)}{L^3}  \sum_{\mathbf{ k} = \frac{2\pi \mathbf{ n}}{L}}^{\mathbf{ n} \in \mathbb{Z}^3 }     \frac{   \tau +  \frac{1}{ 2 \omega_\mathbf{ k} }   }{   (2 \omega_\mathbf{ k})^3 }     e^{-   2 \omega_\mathbf{ k} \tau}     .
   \label{pert2}
\end{align}
The perturbation calculation result  of $  \triangle C^{(2\phi)}( t)  $ given by Eq.(\ref{perturbfinitevolume}) vs. its infinite volume limit is demonstrated in Fig.~\ref{diffCtLpertplot}.

\bibliography{ALL-REF.bib}

\end{document}